\documentclass[11pt,a4paper]{article}
\usepackage{times}
\usepackage{dsfont}
\usepackage{fullpage}
\usepackage{titlesec}
\usepackage[bookmarks=false]{hyperref}
\usepackage{tabulary,multicol,multirow,hhline}

\usepackage[normalem]{ulem}
\usepackage{graphicx,cite}
\usepackage{geometry}
\usepackage{amssymb,amsmath,amsthm}
\usepackage{amssymb,bm}
\usepackage{array}
\usepackage{amsfonts}
\usepackage{enumitem}
\usepackage{color}

\usepackage{geometry,array,colortbl,xcolor}
\usepackage{tabulary}
\usepackage{xcolor}
\usepackage{colortbl}
\usepackage[normalem]{ulem} 
\usepackage{multirow}

\usepackage{bbding}
\usepackage{pifont}
\usepackage{wasysym}
\usepackage{amssymb}
\usepackage{subcaption}
\usepackage{ifthen}

\usepackage{setspace}
\usepackage{amsmath,amsfonts}
\usepackage{amssymb}
\usepackage{graphicx}

\usepackage{amsmath,amscd,amssymb,amsgen,amsfonts,amsbsy,amsthm}
\usepackage{mathrsfs}
\usepackage{bm}
\usepackage{bbm}
\usepackage[utf8x]{inputenc}
\usepackage{color}
\usepackage{textcomp}
\usepackage{float}
\usepackage{latexsym,graphicx}
\usepackage{url}
\usepackage{todonotes}
\usepackage{authblk}

\theoremstyle{plain}
\newtheorem{definition}{Definition}

\newtheorem{theorem}{Theorem}
\newtheorem{lemma}[theorem]{Lemma}
\newtheorem{corollary}[theorem]{Corollary}
\newtheorem{proposition}[theorem]{Proposition}
\newtheorem{conjecture}[theorem]{Conjecture}

\theoremstyle{remark}

\newtheorem{example}{Example}

\makeatletter
\def\highlight#1{%
	\fboxrule1pt %
	\hsize=\dimexpr\hsize-3\fboxrule-3\fboxsep\relax
	\@endpbox\unskip\setbox\lastbox\bgroup
	\fboxrule1pt %
	\fcolorbox{black}{white}{\box{ #1}}\hfill}

\def\Cline#1#2{\@Cline#1#2\@nil}
\def\@Cline#1-#2#3\@nil{%
	\omit
	\@multicnt#1%
	\advance\@multispan\m@ne
	\ifnum\@multicnt=\@ne\@firstofone{&\omit}\fi
	\@multicnt#2%
	\advance\@multicnt-#1%
	\advance\@multispan\@ne
	\leaders\hrule\@height#3\hfill
	\cr}

\begin{document}
\vspace*{3ex}
\begin{center}
	\LARGE{Efficient scheduling in redundancy systems with general service times}
	\vspace*{3ex}
	
	\large
	Elene Anton$^{1}$, Rhonda Righter$^2$ and Ina Maria Verloop$^{3,4}$
\end{center}
\vspace*{1ex}

\begin{center}
	\small
	
	\textsuperscript{1} Université de Pau et des Pays de l’Adour, E2S UPPA, LIUPPA, Anglet, France\\
	\textsuperscript{2} University of California at Berkeley, Berkeley, USA\\
	\textsuperscript{3} CNRS, IRIT, Toulouse, France\\
	\textsuperscript{4} {Universit\'e de Toulouse INP, Toulouse, France}\\
	\normalsize
\end{center}
\vspace*{1ex}

\begin{abstract}We characterize the impact of scheduling policies on the mean response time in nested systems with cancel-on-complete redundancy. We consider not only {class and state}-oblivious policies, such as FCFS and ROS, but also {class-based, and in particular, }redundancy-aware policies of the form $\Pi_1-\Pi_2$, where $\Pi_1$ discriminates among job classes {based on their degree of redundancy} (e.g., least-redundant-first (LRF), most-redundant-first (MRF)) and $\Pi_2$ discriminates among jobs of the same class. Assuming that jobs have independent and identically distributed (i.i.d.) copies we prove the following: (i) When jobs have exponential service times, LRF policies outperform any other policy. (ii) When service times are New-Worse-than-Used, MRF-FCFS outperforms LRF-FCFS as the variability of the service time grows infinitely large. (iii) When service times are New-Better-than-Used, LRF-ROS (resp. MRF-ROS) outperforms LRF-FCFS (resp. MRF-FCFS) in a two-server system. Statement (iii) also holds when job sizes follow a general distribution and have identical copies (all the copies of a job have the same size). Moreover, we show via simulation that, for a large class of redundancy systems, class-based (and, in particular, redundancy-aware) policies can considerably improve the mean response time compared to policies {that ignore the class}. We also explore the effect of redundancy on the stability region.\end{abstract}

\noindent\textit{Key words:} scheduling, redundancy, performance. 


\section{Introduction}

Both 
	empirical (\cite{Ananthanarayanan13,Ananthanarayanan2012,Dean13,Vulimiri13}) and theoretical (\cite{Gardner16,Gardner17b,Joshi15,Lee17a,Anton2019,Shah16}) evidence shows that adding redundancy to parallel-server systems can improve the performance in real-world applications. Under redundancy, an arriving job dispatches multiple copies to all compatible servers, and departs either when a first copy enters service (known as the cancel on start, $c.o.s.$, model) or when a first copy completes service (known as the cancel on complete, $c.o.c.$ model). We focus on $c.o.c.$ models. Redundancy aims to exploit the variability of the queue lengths and server capacities, potentially reducing the response time. However, adding redundant copies also may waste resources in the additional
	servers that do not complete the copy. Hence, the potential of redundancy relies on finding scheduling policies that improve the latency of jobs while not overloading the system. 

In the present paper we investigate the impact of the scheduling policy on the performance of redundancy systems when the usual exponentially distributed i.i.d. copies assumption is relaxed.
In particular, we investigate the performance, in terms of the total number of jobs in the system, for two types of scheduling policies: class-oblivious policies and class-aware policies. Class-oblivious or class-unaware policies are for instance FCFS (First-Come-First-Serve) and ROS (Random-Order-of-Service), where the server ignores the job class. In contrast, under class-aware policies, and in particular, redundancy-aware policies, the scheduler tries to exploit the knowledge of the number of redundant copies, i.e., the degree of a class in the compatibility graph. We generally consider nested systems, in which for any server, all classes with which it is compatible have unique degrees, or levels of redundancy, so our class-aware policies will be redundancy-aware policies.

We propose redundancy-aware policies that are composed of two levels. The first level ($\Pi_1$) describes the priority among the job classes and the second level ($\Pi_2$) describes how the jobs with the same priority are served in a server. Examples of first-level policies are LRF (Least-Redundant-First) and MRF (Most-Redundant-First), where under LRF, respectively MRF, within a server jobs with fewer copies, respectively more copies, have priority over jobs with more copies, respectively fewer copies.  Second-level policies are position-based policies for prioritizing jobs within a class, and could be, e.g., FCFS or ROS. These policies do not depend on the system state, so are easily implementable and scalable.

\subsection{Related work}

Stability of redundancy models has been studied in recent work for exponential service times. Anton et al.\ \cite{Anton2021} provide an overview of the stability results for redundancy models. Under the FCFS scheduling policy and when jobs have independent and identically distributed (i.i.d.) copies, Gardner et al. \cite{Gardner16, Gardner17} and Bonald and Comte \cite{Bonald17a}  fully characterize the stability region and show that it is not reduced due to adding redundant copies. Furthermore, the authors show that the stationary distribution of this model is of a product-form. More precisely, this model falls in the framework of more general systems described in Gardner and Righter \cite{Gardner2020} that present a product-form steady-state distribution. 
Anton et al. \cite{Anton2019} show that the stability condition for the redundancy-$d$ model under FCFS also holds when either PS or ROS is implemented in the servers.

Motivated by the evidence in Vulimiri et al.\ \cite{Vulimiri13} that the i.i.d.\ copies assumption might be unrealistic, Anton et al.\ \cite{Anton2020,Anton2019} assume that copies are identical, that is, all the copies of a job have the same size as the original job. For exponential service times, the authors observe that the stability condition strongly depends on the scheduling policy implemented in the servers. In particular, the stability region of the redundancy-$d$ model is not reduced when the scheduling policy is ROS, but it is dramatically reduced when the scheduling policy is either FCFS or PS.

Raaijmakers et al.\ \cite{Raaijmakers2020} relax the exponential service time assumption to consider New-Better-than-Used (NBU) or New-Worse-than-Used (NWU) service times. They study the redundancy-$d$ model under FCFS, where jobs have identical copies, server capacities are heterogeneous and all servers sample independent speed variations for each copy in service from a general distribution.
That is, the processing time of a copy of a job on any server consists of a fixed job size component that is common across all servers and a server speed component that is chosen independently across servers.
The authors show that under NBU job size distributions,
the  stability region under $d=1$ is larger than that under $d>1$. 
When the service time distribution is instead NWU, the authors observe that the stability condition strongly depends on the load in the system. 

The impact of the redundancy policy on the number of jobs in the system was first studied in \cite{Kim2009,KR08}
for the FCFS scheduling policy, where each job can dispatch i.i.d.\ copies to any server in the system. 
Assuming NWU service time distributions, Koole and Righter \cite{KR08} show that full replication stochastically minimizes the number of jobs in the system at any time. In contrast, for NBU service time distributions, Kim et al. \cite{Kim2009} show that no-replication is optimal. 

Redundancy-aware policies were studied in \cite{Akgun2010, Akgun2013,Gardner2017, GHR19}.
Akgun et al. \cite{Akgun2010,Akgun2013} consider a $c.o.s.$ system where each server has dedicated traffic, that is, each server receives jobs of a class that does not send copies to other servers. The authors consider the DCF (Dedicated-Customers-First, which is equivalent to LRF for their model) service policy and analyze the efficiency and fairness for both dedicated and redundant jobs. Gardner et al. \cite{Gardner2017,GHR19} investigate the impact that the implemented scheduling policy has on the performance for nested $c.o.c.$ redundancy models with exponential service times and i.i.d.\ copies. 

	Gardner et al. \cite{Gardner2017} consider the $W$-model and observe that adding redundancy and implementing FCFS is highly effective in reducing the mean response time in the system, even though LRF is optimal. Moreover, LRF fails to be fair to non-redundant jobs. 
	Thus, the authors propose PF (Primaries-First), which achieves low mean response time while also guaranteeing that a fairness condition is met, and hence is an improvement over FCFS. 
	Nageswaran et al.\ \cite{Nageswaran} also consider fairness, for the $N$-model under FCFS and exponentially distributed job sizes and i.i.d.\ copies where the redundant jobs are scheduled either under $c.o.c.$ or $c.o.s$. The authors analyze the mean response time per class and characterize the conditions under which redundancy is fair compared to the JSQ system without redundancy.
	
	Gardner et al. \cite{GHR19} consider general nested systems and show that for LRF, even though having more redundant jobs is better, the maximum gain comes from adding only a small proportion of redundant jobs. 
	
	\subsection{Contributions}
	
	In this paper, we compare the performance under different redundancy-aware policies for general service time distributions and for both i.i.d.\ and identical copies.  We introduce two-level priority policies $\Pi_1$-$\Pi_2$, where the first level policy $\Pi_1$ determines the priorities among the job classes, and the second-level policy $\Pi_2$ determines the policy among the jobs with the same priority.  
	Below we describe our main contributions, which are summarized in Tables 1 and 2. We note that some of the results hold for a special case of the nested structure that we call the ``Extended W.'' In this model, for each server there is a class that can only be served by that server, and in addition there is a fully redundant class that can be served by all servers.
	
	\begin{table}[b!]
\begin{center}
	\noindent
	\caption{Summary of policy comparison results for i.i.d. copies. We write $\pi\succ\pi'$, if policy $\pi$ has better performance than policy $\pi'$ with respect to the particular performance measure.\\
	}
	\begin{tabular}{|c|c|c|}
		\hline
		\multicolumn{3}{|c|}{\textbf{I.I.D. copies}}\\
		\hline
		\hline
		{\phantom{a} \textbf{Service times} \phantom{a}} &  { \phantom{aa} \textbf{Model} \phantom{aa}} &{\textbf{Results}} \\
		\hline
		Exponential & Nested & \phantom{a} LRF-$\Pi_2 \succ \pi$\phantom{a} (Prop.~\ref{prop:lrf_mrf})  \\
		\hline
		\multirow{2}{*}{\phantom{a} NWU distributions \phantom{a}} & General & $\Pi_1$-FCFS $\succ$ $\Pi_1$-$\Pi_2$ (Prop.~\ref{lem:NWUiid}) \\
		\hhline{|~|-|-|}
		& Extended W &  \phantom{a} MRF-FCFS $\succ$ LRF-FCFS  (Prop.~\ref{prop:Wiid})$^1$\\
		\hline
		\multirow{1}{*}{\phantom{a} NBU distributions \phantom{a}} 
		& W Model & $\Pi_1$-ROS $\succ $ $\Pi_1$-FCFS (Corol.~\ref{prop:WNBUiid})$^2$  \\
		\hline
		\hline 
		{\phantom{a} \textbf{Service times} \phantom{a}} & { \phantom{aa} \textbf{Model} \phantom{aa}} &{\textbf{Conjectures}} \\
		\hline
		\multirow{1}{*}{\phantom{a} NWU distributions \phantom{a}} & General &  \phantom{a} FCFS $\succ \Pi_0$   (Conjecture~\ref{conj:FCFSP0})\\
		\hline
		\multirow{2}{*}{\phantom{a} NBU distributions \phantom{a}} &\multirow{1}{*}{\phantom{a} General \phantom{a}}& $\Pi_1$-ROS $\succ \Pi_1$-FCFS (Conjecture~\ref{conj:ros-fcfs}) \\
		\hhline{|~|-|-|}
		& \multirow{1}{*}{\phantom{a} Nested \phantom{a}} & LRF-$\Pi_2$ $\succ$ MRF-$\Pi_2$ (Conjecture~\ref{conj:lrf-mrf-nbuiid})\\
		\hline
	\end{tabular}
	\label{table}
\end{center}
\footnotesize{Notes: $^1$ as $q\to0$, where the variability of the service time increases as $q\to0$. $^2$ the optimal $\Pi_1$ may idle.
} 
\end{table} 

\begin{table}[b!]
\begin{center}
	\noindent
	\caption{Summary of policy comparison results for identical copies. We write $\pi\succ\pi'$, if policy $\pi$ has a better performance than policy $\pi'$ with respect to the particular performance measure.\\
	}
	\renewcommand{\arraystretch}{1.2}
	\begin{tabular}{|c|c|}
		\hline
		\multicolumn{2}{|c|}{\textbf{Identical copies and general service times}}\\
		\hhline{==}
		{\phantom{aa}\textbf{Model} \phantom{aa}} &{\textbf{Results} }  \\
		\hhline{--}
		\multirow{1}{*}{Extended W} & \phantom{aa} LRF-FCFS $\succ$ MRF-FCFS (Prop.~\ref{prop311})$^3$ \phantom{aa}\\
		\hhline{|-|-|}
		\multirow{1}{*}{Nested} & \phantom{aa}{MRF-$\Pi_2 \succ$ MRF-FCFS (Prop.~\ref{prop:n_ident})}\phantom{aa} \\
		\hline
		W model &  $\Pi_1$-ROS $\succ \Pi_1$-FCFS (Corol.~\ref{prop:WNBUic})$^4$\\ 
		\hline
		\hline
		{\textbf{Model} } &{\textbf{Conjectures} }  \\
		\hhline{--}
		General & $\Pi_1$-ROS $\succ$ $\Pi_1$-FCFS (Conjecture~\ref{conj:ros-fcfs_identical})\\
		\hhline{|-|-|}
		Nested & LRF-$\Pi_2$ $\succ$ MRF-$\Pi_2$ (Conjecture~\ref{conj:lrf-mrf-identica})\\
		\hline
	\end{tabular}
	\label{table}
\end{center}
\footnotesize{Notes: $^3$ with small enough $\lambda$. $^4$ the optimal $\Pi_1$ may idle. 
} 
\end{table} 

Under the i.i.d.~copies assumption, we show that for the nested redundancy model with exponential service times, LRF-$\Pi_2$ minimizes the number of jobs in the system regardless of the non-idling second-level policy $\Pi_2$. That is, we generalize the result in \cite{Gardner2017} to any non-idling second-level policy $\Pi_2$. Furthermore, when service times are NWU, we show that for a given non-idling first-level policy $\Pi_1$, $\Pi_1$-FCFS minimizes the number of jobs in the system, even in non-nested systems. The intuition for the latter comes from the fact that NWU service times are more variable than exponential, so with i.i.d.~copies, the service time of a copy that enters service is stochastically smaller than the remaining service time of a copy of that job that is already in service on another server, which increases the chance that the job departs sooner. Thus, there is a benefit to having multiple copies in service at the same time, which will be more likely under FCFS. We conjecture that FCFS will also be the best single-level policy when service times are NWU.
We further prove that the optimal first-level policy under NWU service times depends on the variability in the service times. In particular, we show, for nested systems with FCFS as the second-level policy, that as the coefficient of variation grows large, MRF becomes optimal, while LRF is optimal when the coefficient of variation is one (exponential service times).  Note that MRF with FCFS maximizes the number of copies served in parallel.

The case of NBU service times is more difficult, and only partial characterizations of an optimal policy are derived for two servers. In this case, it may be optimal to idle (because the second-level policy is non-preemptive), and we are not able to completely characterize conditions under which optimal to idle. However, whenever the server doesn't idle, 
we show that ROS is a better second-level policy than FCFS. This is intuitively clear, since the service time of a copy that is already in service is stochastically smaller than a copy that enters service when service times are NBU, so there is an incentive to minimize the number of copies in service at the same time. For this reason, we also conjecture that ROS will be better than FCFS more generally.

Under the identical copies assumption, we show that LRF is the best first-level policy when the arrival rate is small enough. In addition, we prove that for the nested redundancy model, MRF-$\Pi_2$ outperforms MRF-FCFS for any service time distributions, with $\Pi_2$ non-idling. The latter follows intuitively from the fact that when copies are identical, all the copies of each job have the same size, which necessarily induces a waste of resources when serving copies of the same job. For the $W$-model, our results are similar to those obtained for NBU distributions with i.i.d. copies. This similarity is explained by the fact that having identical copies means that a copy in service has always a smaller remaining service time than a copy of this job that is not yet in service, just as in the NBU i.i.d. case. 

We also considered the impact of having identical copies rather than i.i.d. copies for a fixed policy (Section~\ref{perf:cc}). 
Intuitively, one expects that identical copies will be worse under the $c.o.c.$ discipline, because service on copies started later is more clearly ``wasted.'' When copies are i.i.d, there is a chance that a later started copy will finish earlier. That identical copies is worse in terms of stability under FCFS for the redundancy-$d$ model has been shown in \cite{Anton2019}.
We show that for a general redundancy model with general service times under FCFS or under a two-level policy $\Pi_1$-FCFS for any $\Pi_1$, the total number of jobs when copies are i.i.d. is not only smaller in mean than when copies are identical, the total job process is stochastically smaller pathwise with i.i.d. copies.
Generally, analysis is easier with i.i.d copies, so our result allows for bounds for systems with identical copies.

We also investigate the stability condition for the redundancy-aware scheduling policies analyzed in this paper. In particular, for the redundancy system with a general topology and heterogeneous server capacities and exponentially distributed service times, we show that $(i)$ for LRF-$\Pi_2$ with i.i.d.~copies and non-idling $\Pi_2$, and $(ii)$ for LRF-ROS with any correlation structure among the copies, the stability region is not reduced due to adding redundant copies. 

Finally, we numerically compare redundancy-aware policies with redundancy-oblivious policies. We observe that when service variability is high and copies are i.i.d. or when copies are identical, it may be advantageous to use redundancy-aware policies. 

Tables 1 and 2 summarize our results and related results in the literature. We also include conjectures based on our numerical analysis.

\begin{table}[h!]
\centering
\noindent
\renewcommand{\arraystretch}{1.1}
\caption{Summary of stability results for $c.o.c.$ redundancy models with exponential service times.
}
\begin{tabular}{|c||c|c|c|}
	\hline
	& \multicolumn{1}{c|}{\textbf{\phantom{a}redundancy-$d$\phantom{a}}} &  \multicolumn{1}{c|}{\phantom{a}\textbf{nested}\phantom{a}}&  \multicolumn{1}{c|}{\textbf{\phantom{a}general topology\phantom{a}}} \\
	\hline
	\hline
	\multirow{1}{*}{\textbf{\phantom{a}i.i.d. copies\phantom{a}}} & PS, ROS \cite{Anton2019} & LRF-$\Pi_2$ (Prop.~\ref{stab:lrf} ) & FCFS \cite{Bonald17a,Gardner16} \\
	\hhline{----}
	\multirow{1}{*}{\textbf{\phantom{a}identical copies\phantom{a}}} & FCFS, ROS \cite{Anton2019} & & PS \cite{Anton2021} \\
	\hhline{----}
	\multirow{1}{*}{\textbf{\phantom{a}general copies\phantom{a}}} &   &  LRF-ROS (Prop.~\ref{stab:lrf-ros})   & \\
	\hline
\end{tabular}

\end{table}

\section{Model description}

We consider a $K$ parallel server system with heterogeneous capacities $\mu_s$, for $s\in S$, where $S=\{1,\ldots,K\}$ is the set of all servers. Jobs arrive to the system according to a Poisson process of rate $\lambda$. Each job is labelled with a class~$c$ that represents the subset of servers to which it sends a copy: i.e., $c=\{s_1,\ldots,s_n\}\subset S$, for some $n$. 
We denote by $\mathcal C$ the set of all classes in the system. 
An arriving job is with probability $p_c$ of class~$c$, with $\sum_{c\in\mathcal C}p_c=1$. 
Let us denote by $\mathcal C(s) = \{c\in\mathcal C \ : \ s\in c\}$ the subset of classes that dispatch a copy to server $s$. Therefore, an arrival sends a copy to server~$s\in S$ with probability $\sum_{c\in\mathcal C(s)}p_c$. We assume that the correlation structure among the copies is either i.i.d. or identical copies. We also assume that all copies of a job are canceled once any copy of this job completes.

\begin{figure*}[b!]
\centering
\includegraphics[scale=0.75]{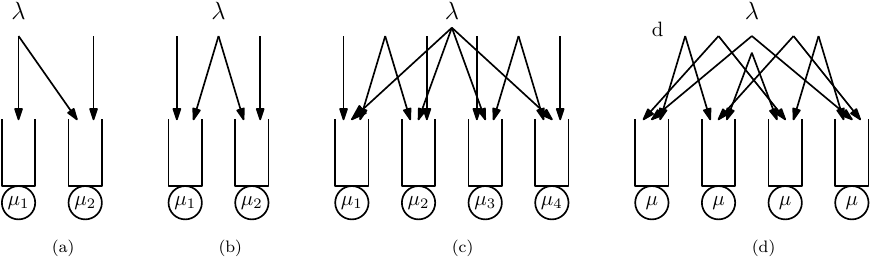}
\caption{(a) the $N$-model, (b) the $W$-model, (c) a general nested model ($K=4$), and (d) the redundancy-$d$ model ($K=4$ and $d=2$).}
\label{fig:red_examples}
\end{figure*}

In this paper, special attention will be given to the class of nested redundancy models.  
We call a redundancy model nested if the set of classes $\mathcal C$ satisfies the following: for all job classes $c,c'\in\mathcal C$, either \emph{i)} $c\subseteq c'$ or \emph{ii)} $c'\subseteq c$ or \emph{iii)} $c\cap c'=\emptyset$.  
The smallest nested system is the so-called $N$-model: this is a $K=2$ server system with classes $\mathcal C=\{\{2\},\{1,2\}\}$. Another nested system is the $W$-model, that is, $K=2$ servers and classes $\mathcal C = \{\{1\}, \{2\}, \{1,2\}\}$. In Figure~\ref{fig:red_examples}, we illustrate   the $N$-model, the $W$-model and a general nested model with $K=4$. 
Nested models arise naturally in systems with data locality constraints, or with hierarchical dispatching.

We also consider the popular redundancy-$d$ model where each incoming job sends a copy to $d$ out of $K$ servers chosen uniformly at random. That is, $\mathcal{C}:=\{ \{s_1,\ldots,s_d\} \subset S  \ : \ s_i\neq s_j, \ \quad \forall i\neq j\},$ with $\vert \mathcal{C} \vert = \binom{K}{d}$ and $p_c=1/\binom{K}{d}$.
We refer to Figure~\ref{fig:red_examples}~(d) for an illustration of a redundancy-$d$ model with $K=4$ and $d=2$.

We denote by $\pi$ a generic scheduling policy implemented in the system. 
We assume that the policy $\pi$ has no information on the actual size of the copies, that is, it is non-anticipating. It may in principle depend on the system state. In this paper, we introduce two-level redundancy-aware (but state-independent) scheduling policies denoted by $\pi=\Pi_1$-$\Pi_2$: 
\begin{itemize}
\item The first-level policy $\Pi_1$ determines the priority among job classes. We assume the priority policy to be strict, that is,  in each server~$s$ there is a strict priority ranking of all job classes in $\mathcal C(s)$. We also assume $\Pi_1$ is preemptive, except in Section 4 on stability where both preemptive and non-preemptive first-level policies are considered.

\item The second-level policy $\Pi_2$ determines the scheduling policy of jobs within the same class at a given server. This policy is assumed to be size-unaware and non-preemptive within the class. That is, once a job of a given class is started at a server, no other job of the same class can be served at that server until the given job has completed (at some server).
\end{itemize}
Examples of first-level policies $\Pi_1$ are Least-Redundant-First (LRF) and Most-Redundant-First (MRF). Note that these policies are uniquely defined for nested systems. Examples of second-level policies $\Pi_2$ are FCFS, LCFS, and ROS.
We note that different servers could adopt different first-level and second-level policies. However, when writing e.g.\ that we consider a system under policy LRF-FCFS, by convention this indicates that \emph{all} servers implement LRF-FCFS.  

We also define class $\Pi_0$ as the single-level redundancy-unaware class of policies that only take into account a job's position in the queue. Again, examples are FCFS, LCFS, and ROS.

For a given scheduling policy $\pi$, we denote by $N^{\pi}_c(t)$ the number of class-$c$ jobs present in the system at time~$t$ and by $N^{\pi}(t):=\sum_{c\in\mathcal C}N^{\pi}_c(t)$ the total number of jobs in the system. We aim to compare the performance of the system, in terms of the number of jobs, under different scheduling policies. We have the following stochastic ordering definition. 

\begin{definition}
For two nonnegative continuous random variables $X$ and $Y$, with respective distributions $F$ and $G$, and $\bar F(x)=1-F(x)$ and $\bar G(x)
=1-G(x)$, we say that $X\geq_{st} Y$, that is, $X$ is stochastically larger than $Y$, if $E[h(X)]\geq E[h(Y)]$ for all increasing functions $h$, or equivalently, if $\bar F(x)\geq \bar G(x)$ for all $x\geq0$.
\end{definition} 

We let $X$ denote the service time distribution of a job when it is served at speed~1. Special focus will be given to exponential service times, as well as  the following two classes of service time distributions: New-Worse-than-Used (NWU) and New-Better-than-Used (NBU), defined below. Let $X_t =\left[X-t \ : \ X>t\right]$ be the remaining processing time of a job that has completed $t$ time units of service.

\begin{definition}
We say that $X$ is New Worse than Used (NWU) if the remaining processing time of a task that has received some processing (is used) is stochastically \emph{larger} than the processing time of a task that has received no processing (is new), i.e., $X_0\leq_{st}X_t$ for all~$t>0$. We say that $X$ is New Better than Used (NBU) if the remaining processing time of a task that has received some processing (is used) is stochastically \emph{smaller} than the processing time of a task that has received no processing (is new), i.e., $X_t\leq_{st}X_0$ for all~$t>0$. (The terminology comes from reliability, in which longer component lifetimes are better.)
\end{definition}   

A sufficient condition for $X$ to be NWU (NBU) is for it to have decreasing (increasing) failure rate, that is, $h(x)=f(x)/\bar F(x)$ is decreasing (increasing) in $x$, where $f(x)$ and $F(x)$ are the probability density function and the cumulative distribution function of $X$, respectively. Additionally, if $X$ has a decreasing (increasing) failure rate, then the coefficient of variation of $X$ is at least (at most) $1$. We note that the exponential service time distribution has constant failure rate, so that it belongs to both categories, NWU and NBU.

Throughout this paper, we provide numerical examples of the performance under the different policies. 
These numerics are obtained  by Matlab where we ran a large number of busy periods $(10^6)$, so that the variance and confidence intervals of the mean number of jobs in the system are sufficiently small. The service time distributions that we consider are exponential, deterministic, and Weibull ($\bar F(x)=e^{(-x/\kappa)^\alpha}$, with $\kappa, \alpha>0$). We note that the deterministic and Weibull distribution with $\alpha\geq1$ are NBU distributions, while the Weibull distribution with $\alpha\leq 1$ is NWU. We also consider a class of degenerate distributions, where $X$ is equal to $Y/q$ with probability $q$ and 0 otherwise, with $Y$ either exponentially distributed or Weibull distributed. Moreover, if $Y$ is a NWU distribution, so is $X$. Throughout this paper we fix $X$ to have unit mean, which in the case of degenerate distributions implies that also $Y$ has unit mean. For the Weibull distribution, we simply fix $\kappa = 1/\Gamma(1/\alpha+1)$ so that the mean equals 1 for any value of $\alpha$.  


\section{Stochastic comparison results}
\label{perf}
In this section, we analyze how the scheduling policy (Section~\ref{perf:sc}) and the copy correlation structure (Section~\ref{perf:cc}) affect the performance of the system.

\subsection{Comparison of scheduling policies}
\label{perf:sc}
For a given system, we compare the total number of jobs with respect to the scheduling policy implemented in the system. We consider i.i.d. copies with exponential, NWU, and NBU service times, and then investigate scheduling policies with identical copies.

\subsubsection{I.I.D. copies and exponential service times}

We first assume that service times are exponentially distributed with i.i.d. copies. Due to the memoryless property, we can show that the number of jobs is insensitive to the implemented second-level policy $\Pi_2$. 
\begin{lemma}
\label{lem:iidpi2}
Consider a redundancy system with a general topology and heterogeneous server capacities, where jobs have exponentially distributed service times and i.i.d. copies. Then, for any $\Pi_2$ and $\Pi'_2$, $\{N^{\Pi_1-\Pi_2}(t)\}_{s\geq0}=_{st} \{N^{\Pi_1-\Pi_2'}(t)\}_{s\geq0}$, where $\Pi_1$ is a strict preemptive priority policy. 
\end{lemma}

\noindent\textit{Proof:}
It is well known that for a single-class queue with i.i.d. exponential service times, and even with time-varying service capacity, the order of service has no effect on the number in queue (though it will effect the variability of waiting times), as long as the service is nonidling. This fact, along with the class-based priority policy $\Pi _{1}$, means that for each server and each class that is compatible with that server, the class will experience time-varying service due to interuptions by higher preemptive priority classes, and service order within the class will not matter.
\qed
\bigskip

\begin{figure}[t!]	\centering
\begin{subfigure}{.49\textwidth}
	\centering
	\includegraphics[width=1.1\textwidth]{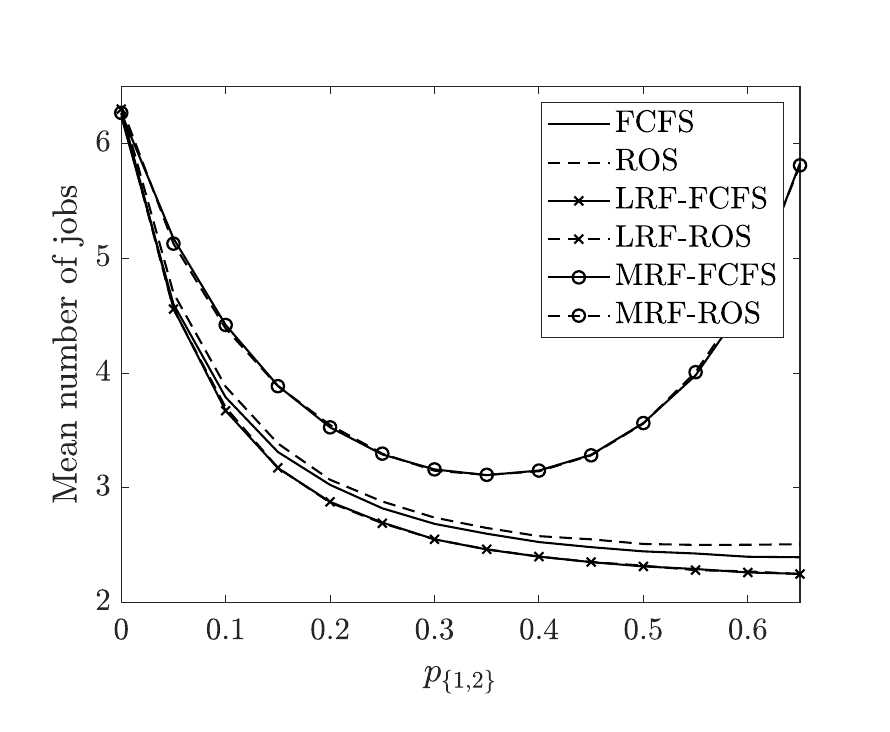}\\
	\caption{$\vec \mu=(1,1)$ and  $\lambda=1.3$}
\end{subfigure}
\begin{subfigure}{.49\textwidth}
	\centering
	\includegraphics[width=1.1\textwidth]{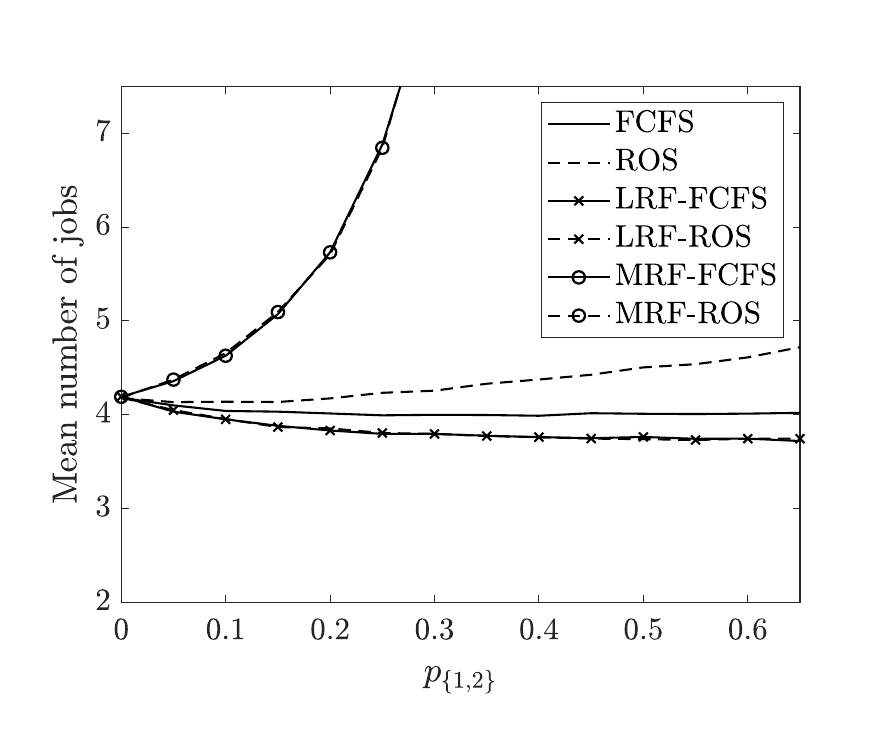}\\
	\caption{$\vec \mu=(1,2)$ and  $\lambda=2$}
\end{subfigure}
\caption{The mean number of jobs for the $W$-model, with $p_{\{1\}}=0.35$ and $p_{\{2\}}=1-p_{\{1\}}-p_{\{1,2\}}$, exponential service times and i.i.d.\ copies.}
\label{fig1}
\end{figure}

We note that the proof goes through under more general assumptions for two-level policies.  Different servers could implement different two-level policies. The top-level policy $\Pi _{1}$ could be state-dependent, and it could decide to serve more than one class at a time by splitting its bandwidth. The second-level policies $\Pi_2$ and $\Pi_2'$ could be either preemptive or nonpreemptive. On the other hand, the result requires a top-level policy $\Pi_1$ that is preemptive and that discriminates among classes.
The latter is illustrated in Figure~\ref{fig1}. There we show that 
a redundancy-oblivious scheduling discipline $\Pi _{0}$ (hence, no strict priority ordering)  does have an impact on the number of jobs in the system. We consider a $W$-model with  $p_{\{1\}}=0.35$ and $p_{\{2\}}=1-p_{\{1\}}-p_{\{1,2\}}$ and vary $p_{\{1,2\}}$. We observe in the figure that  the mean number of jobs under $\Pi_1$-FCFS and $\Pi_1$-ROS coincide for both $\Pi_1$=LRF and, $\Pi_1$=MRF. However, the redundancy-oblivious policies FCFS and ROS  provide different mean numbers of jobs in the system. These policies treat all classes equally, and hence 
Lemma~\ref{lem:iidpi2} does not apply. We observe that FCFS outperforms ROS for any value of $p_{\{1,2\}}$. 
We further note that the differences between FCFS and ROS are most pronounced when the servers are heterogeneous  (Figure~\ref{fig1} (b)) and when $p_{\{1,2\}}$ approaches $1-p_{\{1\}}$.

We also observe in Figure~\ref{fig1} that LRF-$\Pi_2$, with $\Pi_2$=FCFS, ROS, outperform  the other policies. This is consistent with the proposition below, which generalizes the result in~\cite{Gardner2017} for LRF-FCFS.

\begin{proposition}\label{prop:lrf_mrf}
Consider a redundancy system with a nested topology and heterogeneous server capacities where jobs have exponentially distributed i.i.d. copies. Then, 
$$\{N^{LRF-\Pi_2}(t)\}_{t\geq0}\leq_{st} \{N^{\pi}(t)\}_{t\geq0},$$
for any $\Pi_2$ and any $\pi$.
\end{proposition}

\noindent\textit{Proof:} 
In~\cite{Gardner2017} it was proven that 	$\{N^{LRF-FCFS}(t)\}_{t\geq0}\leq_{st} \{N^{\pi}(t)\}_{t\geq0}$ for any policy~$\pi$. Together with Lemma~\ref{lem:iidpi2} this gives the result.  
\qed

\ \\
We note that for non-nested topologies, an optimal policy is expected to be more complex than a two-level policy because an optimal choice of which class to serve will depend on the number of jobs in each class. 
For example, in an $M$-model (job classes $\{1,2\}$ and $\{2,3\}$) with equal arrival rates for all classes and equal service rates at both servers, one would expect that server~2 should serve the longest queue.

\begin{figure}[t!]
\centering
\begin{subfigure}{.49\textwidth}
\centering
\includegraphics[width=1.1\textwidth]{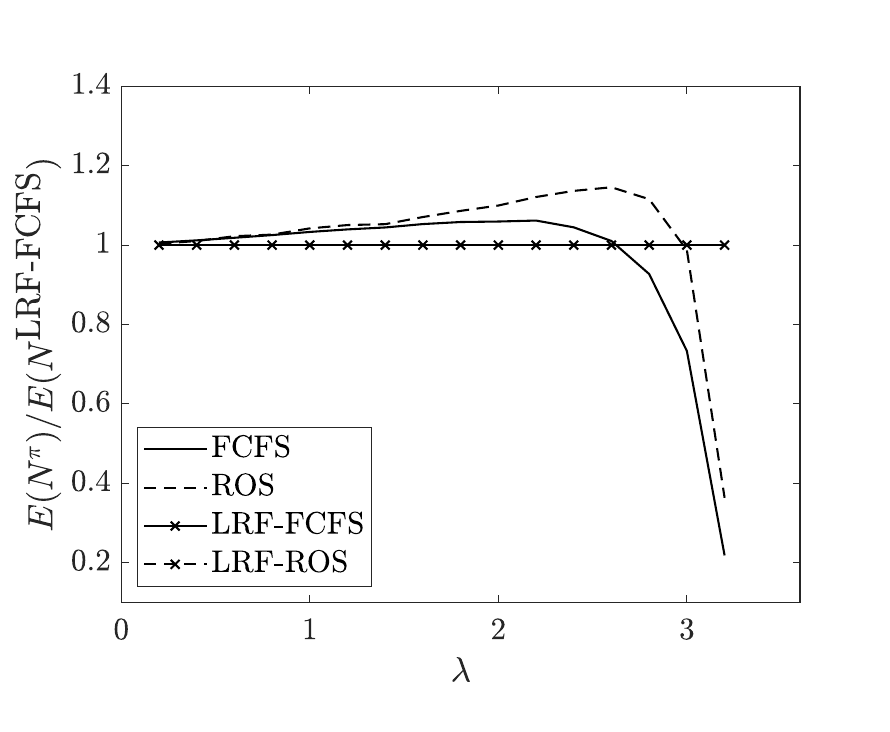}
\end{subfigure}
\caption{The relative mean number of jobs for a non-nested redundancy model with 4 servers, exponential service times and i.i.d.\ copies.}
\label{fig5}
\end{figure}

Figure~\ref{fig5} shows the simulation results for another example with a non-nested topology, but here LRF is still a strict priority first-level policy  (in each server there is at most one class of job with a given number of copies). In particular, there are $K=4$ servers with homogeneous capacities $\vec\mu=(1,1,1,1)$ and a non-nested redundancy topology $\mathcal C=\{\{3\},\{4\},\{1,2\},\{3,4\},\{2,3,4\}\}$, and $\vec{p}=(0.15,0.20,0.15,0.15, 0.35)$. We observe that FCFS outperforms LRF-FCFS when the load in the system is sufficiently large.


\subsubsection{I.I.D.\ copies and NWU service times}

When jobs have i.i.d.\ copies and NWU service time distributions, the service time of a copy that is already in service is stochastically larger than that of an i.i.d.\ copy that has not received service yet. Hence, this suggests that whenever a server becomes available to a class, it will be better to serve a copy of a job that has already a copy elsewhere in service, to have it leave faster.  That is exactly what policy $\Pi_2$=FCFS does. In the result below we show that, given a first-level policy $\Pi_1$,  FCFS  at the second level is indeed optimal. We note that in~\cite{KR08} this result was proved for the redundancy system with only one class of jobs.  The proof is deferred to Appendix~A.

\begin{proposition}
\label{lem:NWUiid}
Consider a redundancy system with a general topology, heterogeneous server capacities, NWU service times and i.i.d. copies.  Then, $$\{N^{\Pi_1-FCFS}(t)\}_{t\geq0}\leq _{st} \{N^{\Pi_1-\Pi_2}(t)\}_{t\geq0},$$ for all $t\geq0$ and any second-level 
policy $\Pi_2$.
\end{proposition}
\begin{figure}[t!]
\centering
\begin{subfigure}{.49\textwidth}
\centering
\includegraphics[width=1.1\textwidth]{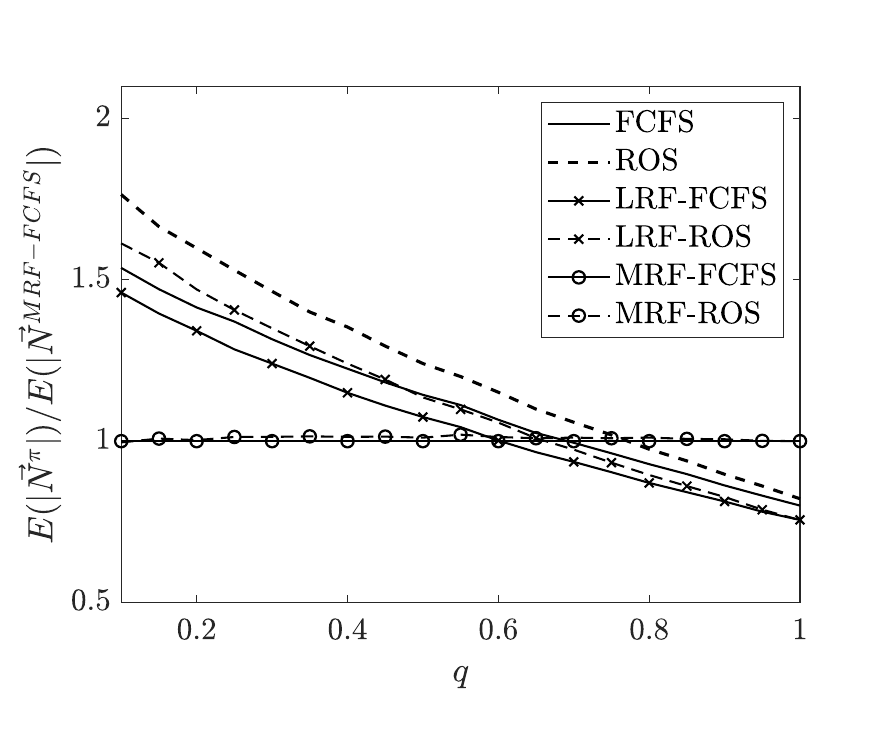}
\caption{$Y\sim$ exponential}
\end{subfigure}
\begin{subfigure}{.49\textwidth}
\centering
\includegraphics[width=1.1\textwidth]{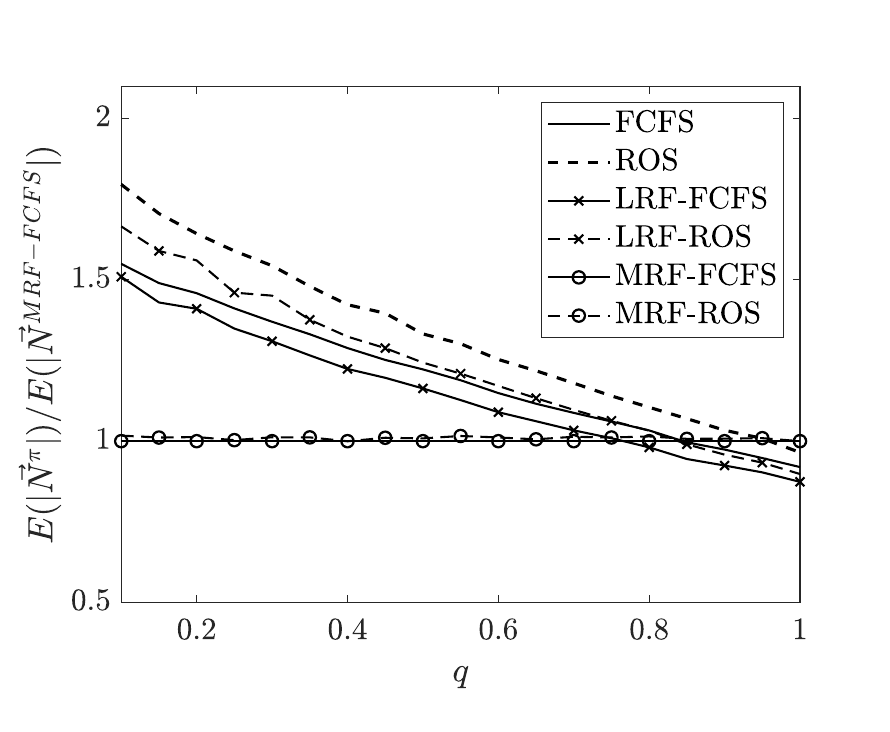}
\caption{$Y\sim$  Weibull: $\alpha=0.75$, $C^2_Y=1.83$}
\end{subfigure}
\begin{subfigure}{.49\textwidth}
\centering
\includegraphics[width=1.1\textwidth]{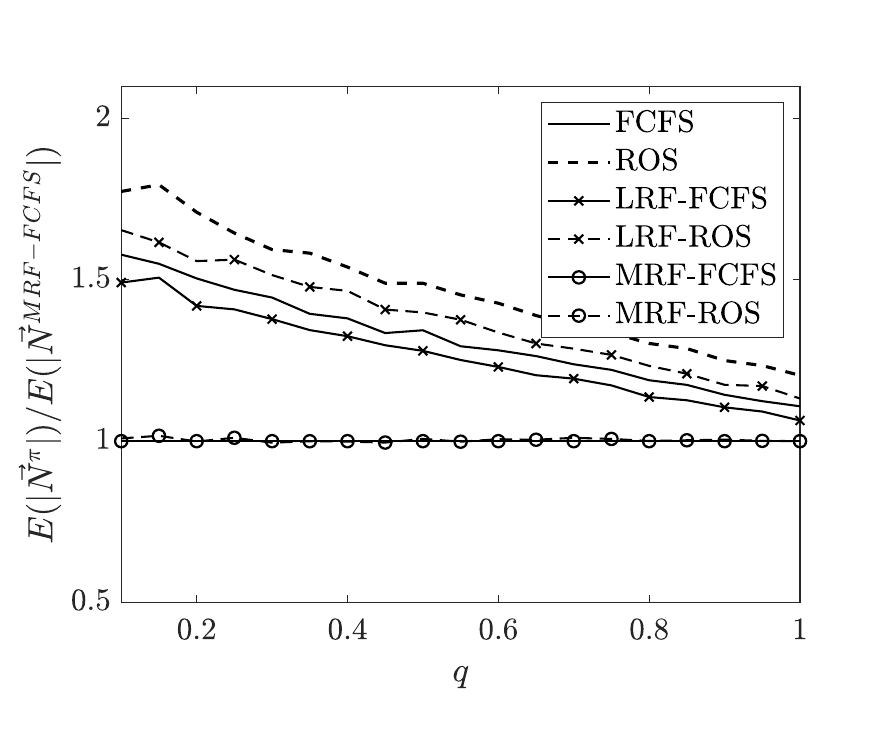}
\caption{$Y\sim$ Weibull: $\alpha=0.5$, $C^2_Y=5$.}
\end{subfigure}

\caption{The mean number of jobs for the $W$-model with i.i.d. copies when $\vec \mu=(1,1)$, $\lambda=1.5$ and $p_c=1/3$ for all $c\in\mathcal C$ with $X$ a mixture of $Y$ (top left) exponential, (top right) Weibull with parameters $\alpha=0.75$ and $\kappa=0.83$, and (bottom) Weibull with parameters $\alpha=0.5$ and $\kappa=0.5$.}
\label{fig2}
\end{figure}

As an illustration, in Figure~\ref{fig2} we simulate the $W$-model with  $\lambda=1.5$, homogeneous capacities $\vec\mu=(1,1)$ and $p_c=1/3$ for all $c\in\mathcal C$.
We assume that the service time distribution $X$ is a mixture of $Y/q$ with probability $q$ and $0$ otherwise, where $Y$ is NWU. In Figure~\ref{fig2}~(a) we chose $Y$ to be exponential. In Figure~\ref{fig2}~(b) and~(c) we chose $Y$ to be Weibull. The squared coefficient of variation of $X$ equals $C^2 = \frac{\mathbb{E}(Y^2)}{q\mathbb E(Y)^2}- 1$ and increases without bound when $q\to0$. We note that in the special case where $Y$ is exponentially distributed, we have $C^2=2/q - 1$. 
Consistent with Proposition~\ref{lem:NWUiid}, we observe that $\Pi_1$-FCFS (solid line) outperforms $\Pi_1$-ROS (dashed line) for both $\Pi_1$=LRF ($\times$) and $\Pi_2$=MRF ($\circ$). This observation also holds for single-level redundancy-oblivious policies, i.e., FCFS outperforms ROS. In fact, we conjecture that the single-level policy FCFS is better than any other single-level policy $\Pi_0$, however, we did not obtain a proof for this.

\begin{conjecture}
\label{conj:FCFSP0}
Consider a redundancy system with a general topology, heterogeneous server capacity, NWU service times and i.i.d.\ copies. We conjecture that FCFS outperforms (in some sense) any other single-level policy~$\Pi_0$.
\end{conjecture}

In Figure~\ref{fig2} (a) and (b) we also observe that as $q$ approaches 1, LRF-FCFS outperforms the other policies. In fact, when $q=1$ and $Y$ is exponentially distributed, it was shown in Proposition~\ref{prop:lrf_mrf} that LRF-FCFS minimizes the number of jobs.
When $q$ approaches 0, that is $C^2\to\infty$, we observe that MRF-FCFS outperforms all other scheduling policies. 
This example shows that there is not a unique first-level policy that minimizes the total number of jobs in the system for non-exponential service times. 

Under NWU service times,  MRF might perform well because this policy serves at all times the copies of the job that has the most copies.
However, given we consider nested systems, LRF is the first-level policy that minimizes the time that a server is idle.  Hence, there is a trade-off, and which policy is optimal will strongly depend on the coefficient of variation of the service time distribution, which impacts how beneficial it is to serve copies of the same job (the more variable services are, the more profitable to serve multiple copies). 
The proposition below supports our observation for a multi-server extended $W$ topology system, where each server has dedicated traffic and there is one flexible class of jobs that sends copies to all servers, that is, $\mathcal C = \{ \{ s\}_{s\in S}, S\}$. 
The proof can be found in Appendix~A.

\begin{proposition}\label{prop:Wiid}
Consider an extended $W$-model with $K$ heterogeneous servers with capacities $\mu_s$ and i.i.d.\ copies.
We assume that the service time distribution $X$ is a mixture of $Y/q$ with probability $q$ and $0$ otherwise, where $Y$ is NWU. 
Then, 
$$q\mathbb{E}(N^{\rm{MRF-FCFS}}) <  q\mathbb{E}(N^{\rm{LRF-FCFS}}) + o(1), \textrm{ as } q\to 0.$$ 
\end{proposition}

Combining Proposition~\ref{prop:Wiid} with Proposition~\ref{lem:NWUiid} implies that for the heterogeneous server system with $\mathcal C = \{ \{ s\}_{s\in S}, S\}$, i.i.d. copies, and $X$ being a mixture of $Y/q$ with probability $q$ and $0$ otherwise, where $Y$ is NWU, the MRF-FCFS policy is better than any other two-level policy as $q\to0$. 

\subsubsection{I.I.D. copies and NBU service times} 

In the case of NBU service times and i.i.d. copies, we are only able to obtain partial results.
This is not surprising, given that already when all
jobs are of the same class, the only known results are for the case of two homogeneous servers \cite{KR08}, or for an arbitrary number of servers but a saturated system \cite{Kim2009}. 
The problem is complicated by the fact that for two-level policies with no preemption at the second level, when service times are not exponential it may be optimal to idle a server.

We first show that for two-level policies, when a server $s$ has a dedicated class $\{s\}$, i.e., a class that can only be served by that server, then it is better to serve a job of class $\{s\}$ rather than idle whenever class $\{s\}$ has priority over other classes currently present at the server. This is in fact true regardless of the service time distribution. We then provide a further partial characterization of the optimal second-level policy for
the two-server nested system, i.e., a $W$-model with NBU service times. 
%
The proofs can be found in Appendix~A.

\begin{proposition}\label{prop:iid_NBU_dedicated_nonidling}
Consider a model with a general topology,  
heterogeneous servers, general service times, i.i.d.\ copies, and a fixed two-level policy $\Pi_1-\Pi_2$. Server $s$ should never idle when there are jobs of class $\{s\}$ present and when that class has priority over all other classes present at server $s$.
\end{proposition}

For the following proposition and its corollary, we assume a $W$-model, so the first-level policy is either LRF or MRF. However, we do not require both servers to follow the same first-level policy.

\begin{proposition}\label{prop:iid_NBU_2_FCFS}
Consider a $W$-model  
with heterogeneous servers, NBU service times and i.i.d.\ copies. 
Whenever the first-level priority policy at either server $s$ gives priority to class $\{1,2\}$ and the second-level policy decides not to idle, it is stochastically optimal to serve a copy of a class-$\{1,2\}$ job that has no copy yet in service in the other server (if possible). 
This is true for any fixed policy for the other server.
\end{proposition}

The proposition above assumes that we know whether a class-$\{1,2\}$ job has 
received service at the other server. In practice, this information might not be available. In the following proposition, we therefore provide a comparison between FCFS and ROS and show that given the decision to serve a class-$\{1,2\}$ job, it is better to choose according to ROS, as this chooses with a higher chance a copy that has not received service elsewhere yet.  The proof can be found in Appendix~A.

\begin{corollary}\label{prop:WNBUiid}
Consider a  $W$-model with heterogeneous servers, NBU service times and i.i.d.\ copies 
following a two-level policy. 
The optimal policy may idle, but whenever a job is served,
it is stochastically better for the second-level policy to serve
according to ROS than FCFS. 
\end{corollary}

\begin{figure}[b!]
\centering
\begin{subfigure}{.49\textwidth}
\centering
\includegraphics[width=1.1\textwidth]{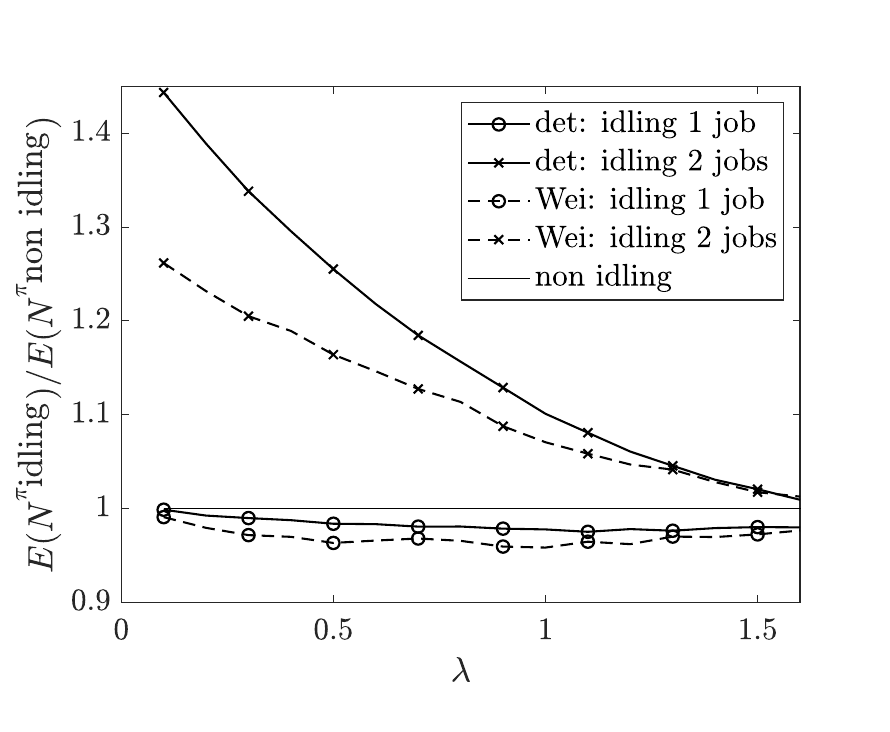}
\subcaption{LRF-ROS}
\end{subfigure}
\begin{subfigure}{.49\textwidth}
\centering
\includegraphics[width=1.1\textwidth]{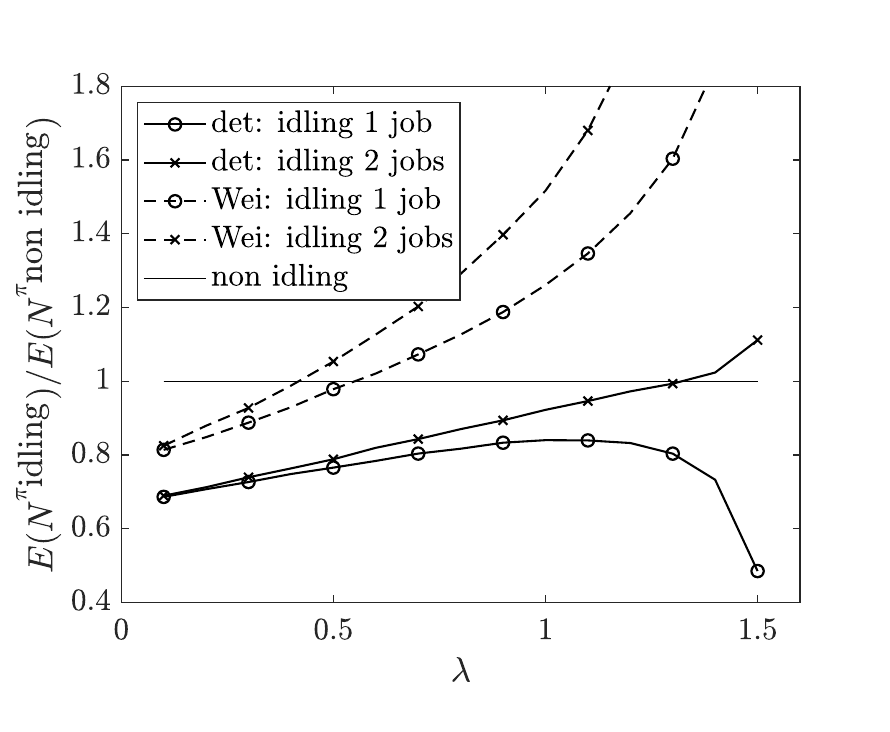}
\subcaption{MRF-ROS}
\end{subfigure}

\caption{The mean number of jobs for the $W$-model with i.i.d. copies with respect to $\lambda$, with capacities $\vec\mu=(1,1)$, $p_c=1/3$, and NBU service times: either deterministic (det, full line) or Weibull with $\alpha=1.25$ (Wei, dashed line).}
\label{fig62}
\end{figure}

We note that the proof of the previous two propositions can not be generalized to more than two servers. The reason for this is given at the end of the proof of Proposition~\ref{prop:iid_NBU_2_FCFS} in Appendix~A. 

The above propositions do not completely characterize the optimal second-level policy. For example, it does not tell us whether or not a server should idle when there is only one class-$\{1,2\}$ job that has already received some service in the other server. 
Idling may be optimal because we do not allow preemption at the second level. For example, under $\Pi_1$-ROS we expect the optimal idling policy to depend on the number of class-$\{1,2\}$ jobs since it might be profitable to idle if and only if there are fewer than $x$ class-$\{1,2\}$ jobs present for some $s$, because the probability of selecting the same class-$\{1,2\}$ job under ROS is decreasing in~$x$.
In Figure~\ref{fig62} we consider the $W$-model and simulate the performance of LRF-ROS (a) and MRF-ROS (b) when server 1 idles when according to the first-level policy it should start serving a new class-$\{1,2\}$ job, but there are only $x$ class-$\{1,2\}$ jobs present, with $x$ equal to either $1$ or $2$. For LRF-ROS (a) we observe that idling when there is only 1 class-$\{1,2\}$ job present is better than non-idling. For MRF-ROS we observe that for small enough arrival rate $\lambda$,  the policies that idle when there are 1 class-$\{1,2\}$ jobs or up to 2 class-$\{1,2\}$ jobs are better than non-idling. However, as $\lambda$ increases the performance of these policies strongly depends on the service time distribution.

\begin{figure}[b!]
\centering
\begin{subfigure}{.49\textwidth}
\centering
\includegraphics[width=1.1\textwidth]{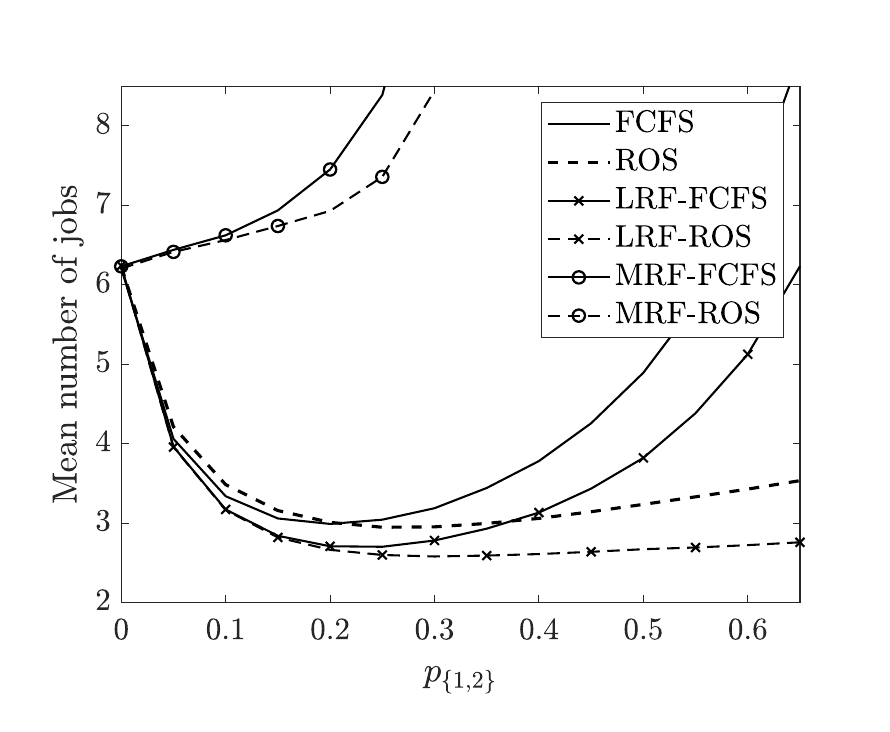}\\
\caption{Deterministic}
\end{subfigure}
\begin{subfigure}{.49\textwidth}
\centering
\includegraphics[width=1.1\textwidth]{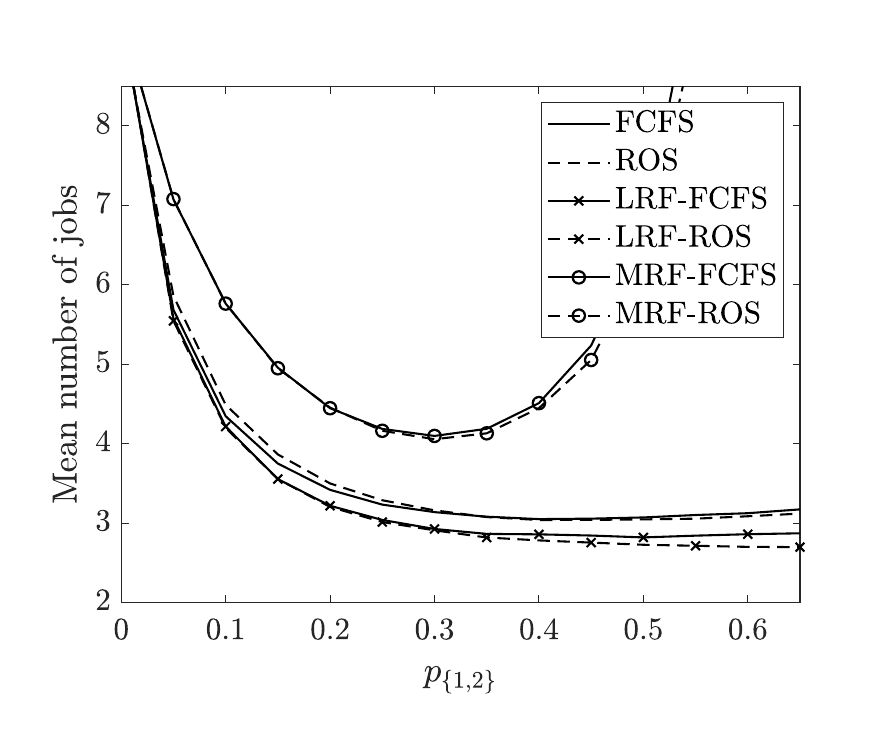}\\
\caption{Weibull with $\alpha=1.25$}
\end{subfigure}
\caption{The mean number of jobs for the $W$-model with i.i.d.~copies and capacities $\vec\mu=(1,1)$, $\lambda=1.4$, $p_{\{1\}}=0.35$ and $p_{\{2\}} = 1-p_{\{1\}}-p_{\{1,2\}}$.}
\label{fig61}
\end{figure}

In general, with NBU service times, we want to avoid serving multiple copies of a job at the same time, because service at one of the servers is likely to be ``wasted.'' Under ROS we are less likely to have multiple copies in service at the same time than under FCFS. This is the intuition for our conjecture below, that ROS is better than FCFS for the second-level policy among two-level policies that are nonidling. However, we were not able to prove the conjecture, because our sample-path proof approach for similar results uses both the option of idling, and the non-optimality of idling, and the latter is not true here.
After the conjecture we provide several numerical results supporting our conjecture.

\begin{conjecture}\label{conj:ros-fcfs}
Consider a redundancy system with a general topology, heterogeneous server capacities, NBU service times and i.i.d.\ copies, and assume idling is not permitted. Then we conjecture that for a given first-level policy~$\Pi_1$, $\Pi_1$-ROS outperforms (in some sense) $\Pi_1$-FCFS.
\end{conjecture}

\begin{figure}[t!]
\centering
\begin{subfigure}{.49\textwidth}
\centering
\includegraphics[width=1.1\textwidth]{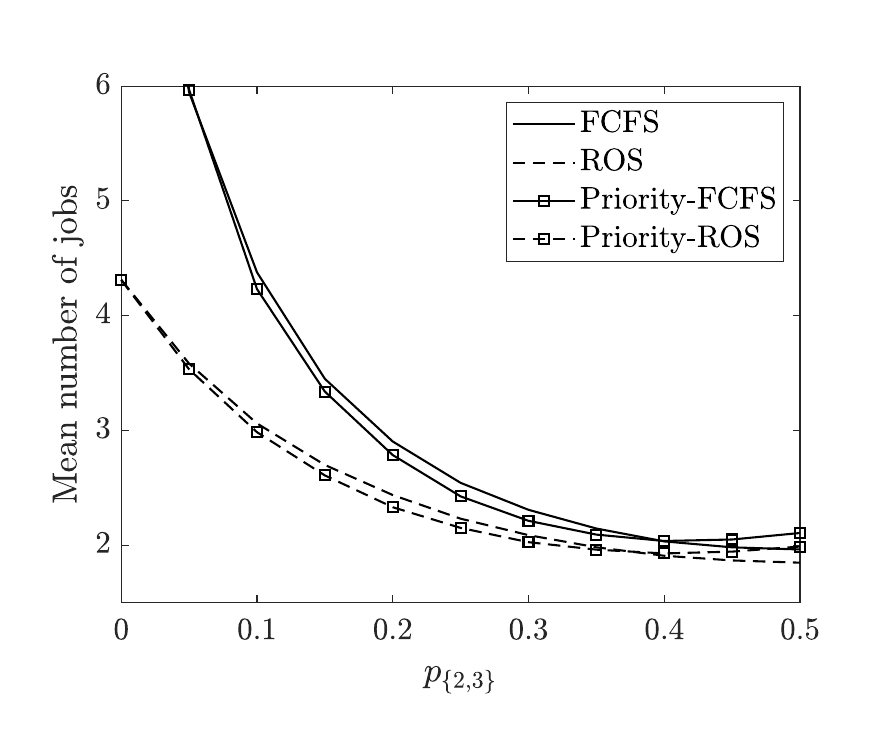}\\
\caption{$M$-model}
\end{subfigure}
\begin{subfigure}{.49\textwidth}
\centering
\includegraphics[width=1.1\textwidth]{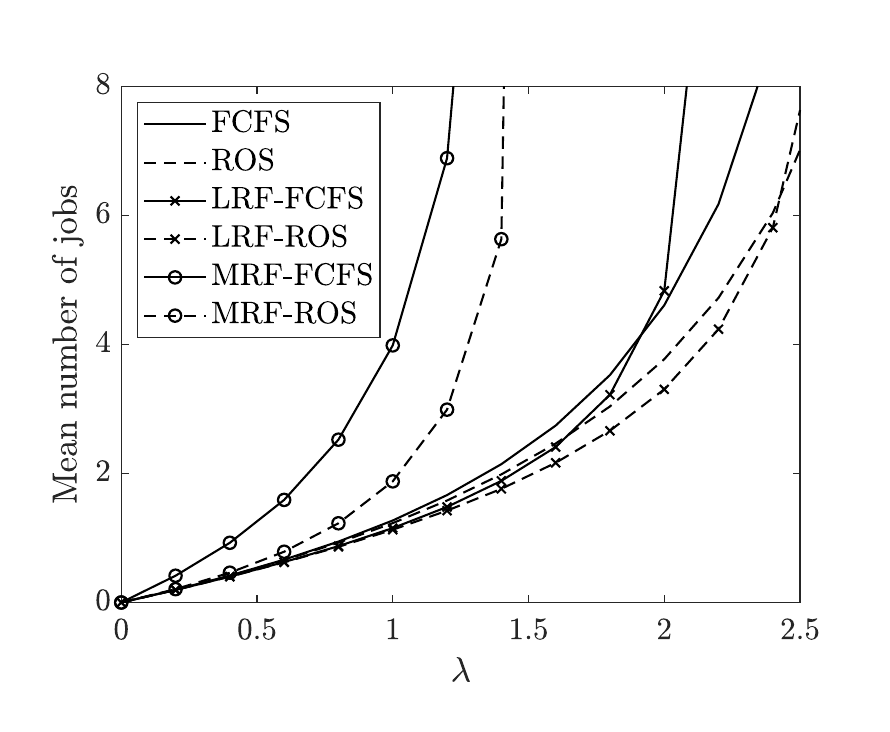}\\
\caption{Non-nested model as in Fig.~\ref{fig5}}
\end{subfigure}
\caption{The mean number of jobs with Weibull service times ((a) $\alpha=1.25$ and (b) $\alpha=10$) and i.i.d.~copies. 
}
\label{fig63}
\end{figure}

In Figure~\ref{fig61}, we consider the $W$-model and 
compare different policies assuming there is no idling, and 
observe that for a given first-level policy, ROS outperforms FCFS (the solid lines are above the dashed lines), consistent with Conjecture~\ref{conj:ros-fcfs}. We note that this is not the case for the single-level policies ROS and FCFS. 
In Figure~\ref{fig63} we then compare ROS and FCFS for \emph{non-nested} models.
In Figure~\ref{fig63} (a) we consider the $M$-model ($K=3$ homogeneous servers with   $\mathcal C=\{\{1,2\},\{2,3\}\}$ and $\lambda=1.6$), and plot the mean number of jobs in the system with respect to $p_{\{2,3\}}$. Note that FCFS is equivalent to LRF-FCFS and MRF-FCFS, and similarly for ROS, because the two job classes both have the same number of copies. We also show results for the Priority-$\Pi_2$ policy, in which jobs of class $\{1,2\}$ have preemptive priority over jobs of class $\{2,3\}$ and among jobs of the same class, copies are served according to $\Pi_2$.
In Figure~\ref{fig63} (b) we consider the same non-nested redundancy model as in Figure~\ref{fig5}, and plot the mean number of jobs in the system with Weibull service times ($\alpha=10$). We observe that for both non-nested models,  $\Pi_1$-ROS outperforms $\Pi_1$-FCFS, which is consistent with Conjecture~\ref{conj:ros-fcfs}.



We do not have a comparison result for the first-level policies. Numerically we did observe though that LRF outperforms MRF for both deterministic and Weibull NBU service times, see Figure~\ref{fig61}. In the case of deterministic service times, we can indeed prove this for a particular redundancy model (it follows directly from Proposition~\ref{prop311} below with deterministic services). For NBU service times and i.i.d. copies, we provide the following conjecture. 
Again, the intuition is that, with NBU service times, we want to avoid serving multiple copies of the same job at the same time, and postponing more redundant jobs will decrease the likelihood of such ``wasted'' service. And again, our proof technique breaks down because idling may be optimal.

\begin{conjecture}
\label{conj:lrf-mrf-nbuiid}
Consider the nested redundancy model, heterogeneous server capacities, NBU service times and i.i.d. copies. We conjecture that for a given second-level scheduling policy $\Pi_2$,  it holds that 
LRF-$\Pi_2$ outperforms (in some sense) MRF-$\Pi_2$.
\end{conjecture}

\subsubsection{Identical copies}
\label{subset:id}
In this section we consider identical copies and general service times and investigate the impact of the first-level and second-level policies.  

\begin{figure}[t!]
\centering
\begin{subfigure}{.49\textwidth}
\centering
\includegraphics[width=1\textwidth]{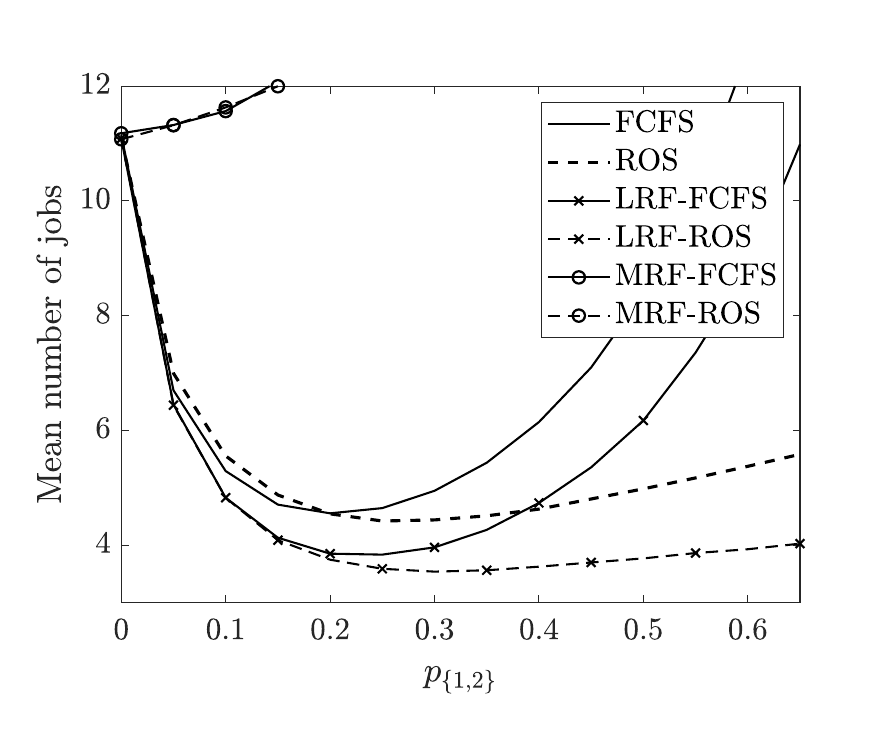}\\
\caption{Exponential, $\lambda=1.4, \vec\mu=(1,1)$}
\end{subfigure}
\begin{subfigure}{.49\textwidth}
\centering
\includegraphics[width=1\textwidth]{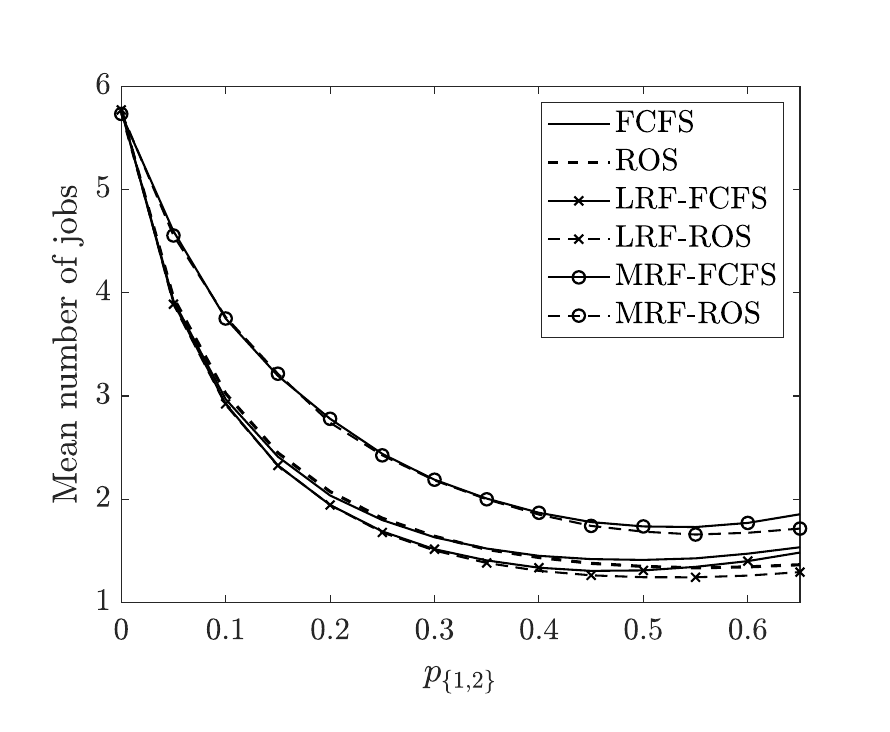}\\
\caption{Exponential, $\lambda=1.3, \vec\mu=(2,1)$}
\end{subfigure}

\begin{subfigure}{.49\textwidth}
\centering
\includegraphics[width=1\textwidth]{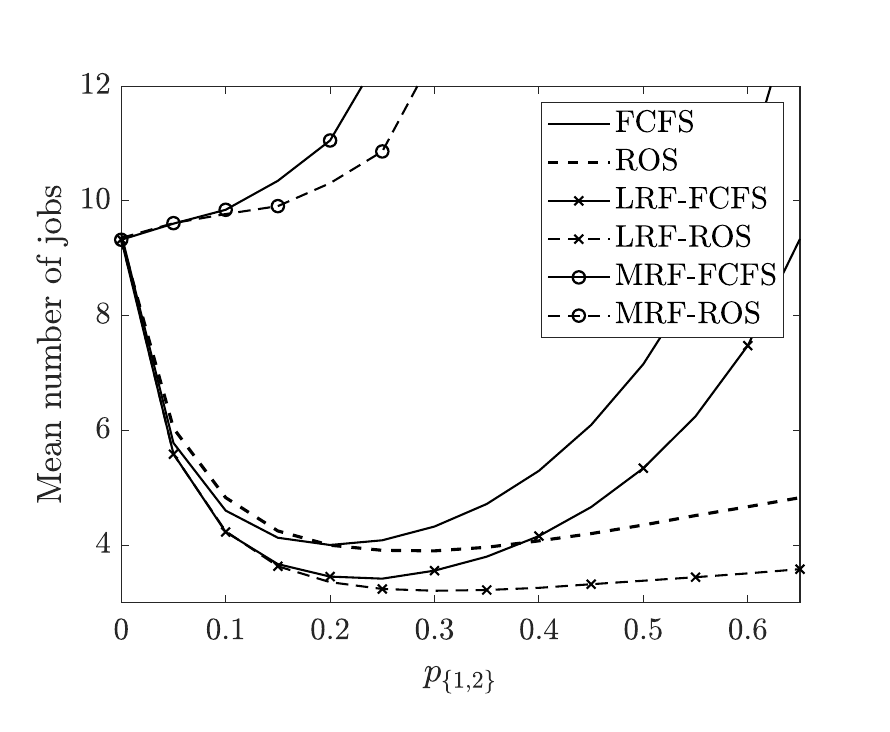}\\
\caption{Weibull, $\alpha=1.25$ (NBU), \\
	$\lambda=1.4, \vec\mu=(1,1)$}
	\end{subfigure}
	\begin{subfigure}{.49\textwidth}
\centering
\includegraphics[width=1\textwidth]{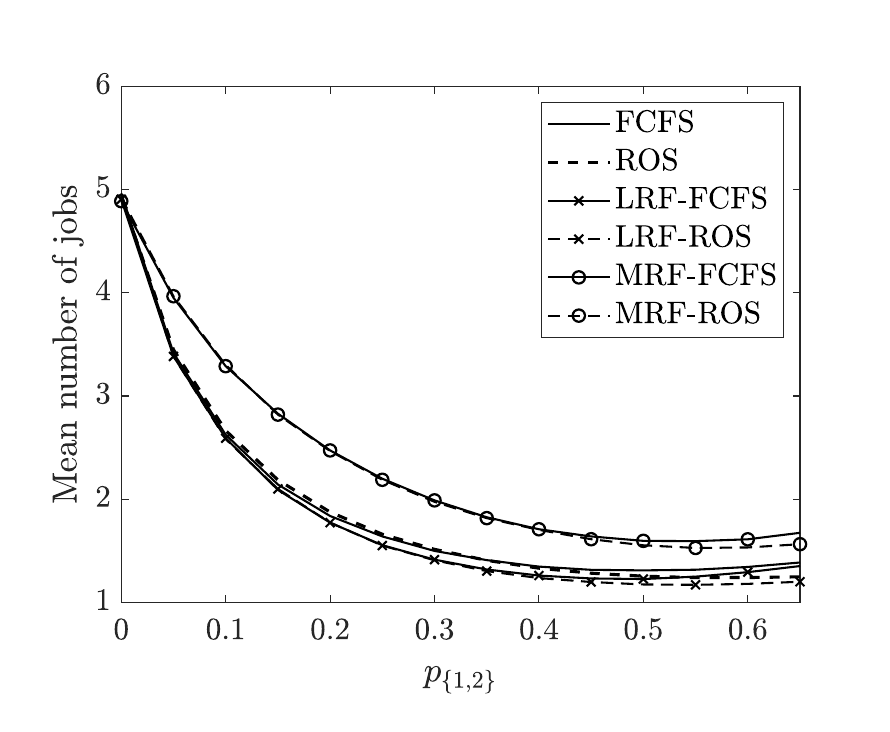}\\
\caption{Weibull, $\alpha=1.25$ (NBU), \\
	$\lambda=1.3, \vec\mu=(2,1)$}
	\end{subfigure}
	
	\begin{subfigure}{.49\textwidth}
\centering
\includegraphics[width=1\textwidth]{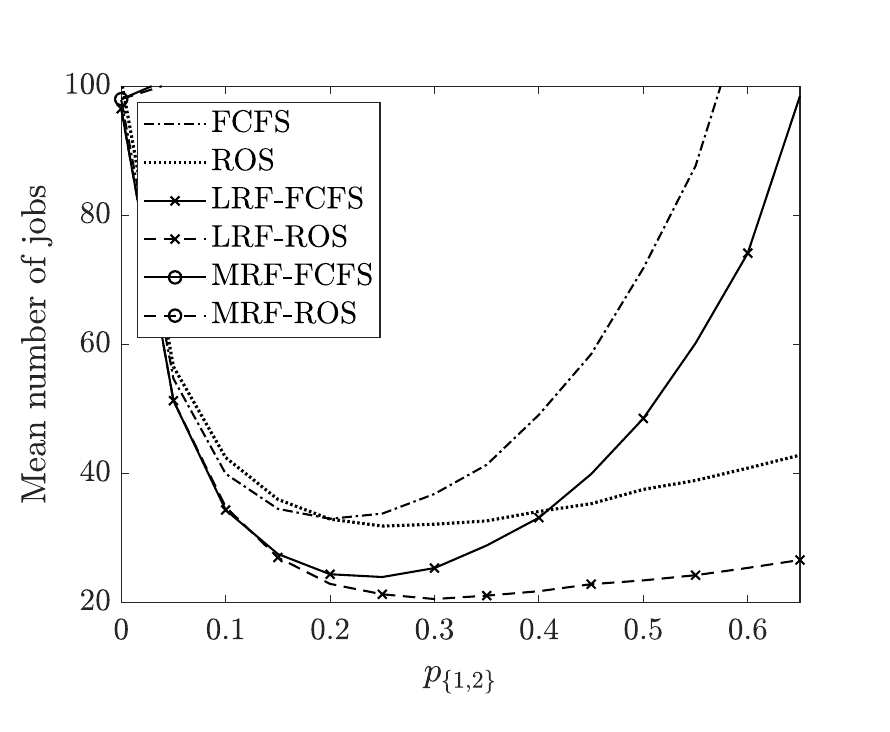}\\
\caption{Mixture exp., $q=0.1$ (NWU), \\
	$\lambda=1.4, \vec\mu=(1,1)$}
	\end{subfigure}
	\begin{subfigure}{.49\textwidth}
\centering
\includegraphics[width=1\textwidth]{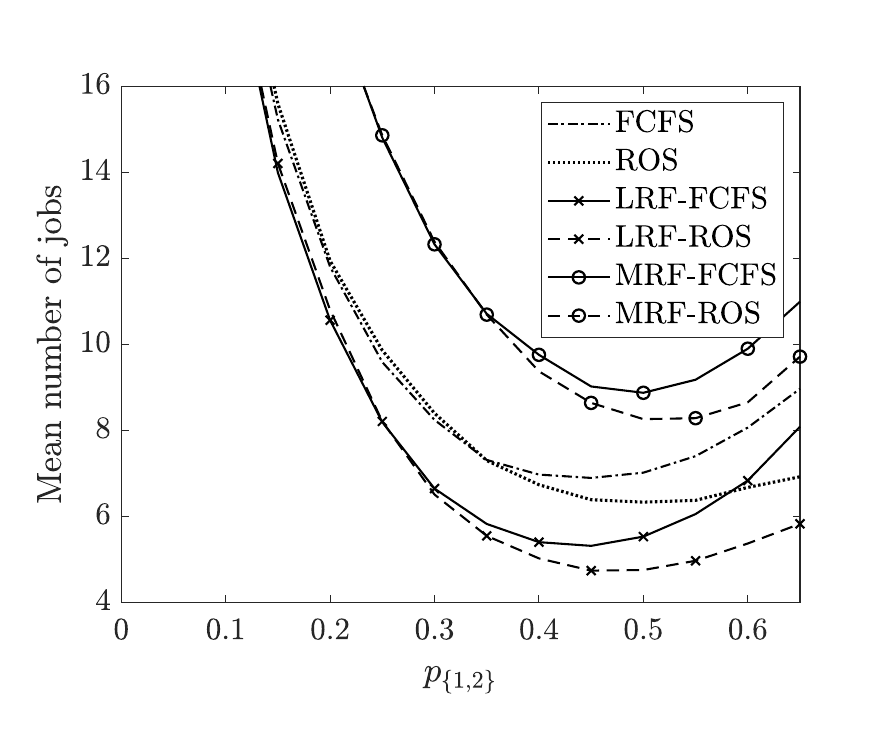}
\caption{Mixture exp., $q=0.1$ (NWU), \\ 
	$\lambda=1.3, \vec\mu=(2,1)$}
	\end{subfigure}
	\caption{The mean number of jobs for the $W$-model with $p_{\{1\}}=0.35$, $p_{\{2\}}=1-p_{\{1\}}-p_{\{1,2\}}$ and identical copies.}
	\label{fig4}
\end{figure}

In Figure~\ref{fig4}, we plot the mean number of jobs under different policies for the $W$-model with identical copies and  several choices of the service time distributions. We observe that LRF outperforms MRF for a given service time distribution and second-level policy. This can be explained as follows. When copies are identical, having several copies of the same job in service implies that capacity of one of the servers is unnecessarily dedicated to this job. Since LRF minimizes the number of copies of the same job in service, one would expect LRF to be optimal.
In the result below, we show that this can be proved  under certain conditions when the second-level policy is FCFS. The proof of the result can be found in Appendix~A and follows by upper bounding the LRF system and using stochastic coupling arguments.

\begin{proposition}
\label{prop311}
Consider the extended W model with $K$ heterogeneous servers with capacities $\mu_s$ where $\mu_1=\max_{s\in S}\{\mu_s\}$ and each server has a dedicated job class and there is an additional job class that sends 
copies to all the servers. That is, $\mathcal C = \{ \{ s\}_{s\in S}, S\}$. We assume general service times and identical copies. Assume that $p_{S} \geq p_{\{1\}}$ or that $\lambda\leq\lambda_0$, where $\lambda_0$ is given by Equation~\eqref{eq:2} in Appendix~A. 
Then it holds that 
$$\mathbb{E}(N^{LRF-FCFS})\leq \mathbb{E}(N^{MRF-FCFS}).
$$
\end{proposition}

For the second-level policy, we observe in Figure~\ref{fig4} that ROS performs better than FCFS. In the case of MRF as the first-level policy, we have a more general result, that is, MRF-FCFS is worse than any other MRF-$\Pi_2$ policy. The proof is deferred to the Appendix~A.

\begin{proposition}\label{prop:n_ident}
Consider a redundancy system with a nested topology  and heterogeneous server capacities, where the service times of the jobs are distributed according to a general distribution and copies are identical. Then,
$\{N^{MRF-FCFS}(t)\}_{t\geq0}\geq_{st} \{N^{MRF-\Pi_2}(t)\}_{t\geq0}$  with preemptive MRF and where $\Pi_2$ is non-idling.
\end{proposition}

The key property to prove the above result is that in a nested redundancy system with  preemptive MRF-FCFS, all copies of a job enter service simultaneously and complete service in the server with highest capacity, wasting resources on the other servers serving this job. Hence, the system behaves as if each job is always served by its compatible server with the highest capacity, while the other compatible servers effectively idle until that job has departed. 

The above result gives the worst second-level policy. It is however less clear what is the best second-level policy. We obtained the following partial characterizations of the optimal second-level policy, mostly restricted to the W model. These partial characterizations coincide with those of Propositions 
\ref{prop:iid_NBU_dedicated_nonidling} and Propositions~\ref{prop:iid_NBU_2_FCFS} and Corollary~\ref{prop:WNBUiid} for i.i.d.\ copies and NBU service times, and their proofs are the same.

\begin{proposition}\label{prop:identical_dedicated_nonidling}
Consider a model with a general topology,  
heterogeneous servers, identical copies, and a fixed first-level policy $\Pi_1$. Server $s$ should never idle when there are jobs of class $\{s\}$ present and when that class has priority over all other classes present at server $s$.
\end{proposition}

\begin{proposition}\label{prop:identical_FCFS}
Consider a $W$-model (so the first-level priority policy is either LRF or MRF)
with heterogeneous servers and identical copies.  Whenever the first-level priority policy gives priority to class-$\{1,2\}$ and the second-level policy decides not to idle, it is stochastically optimal to choose to serve a copy of a class-$\{1,2\}$ job that has no copy yet in service in the other server (if possible). Server $s$ should never idle when class $\{s\}$ has priority and there are jobs of class $\{s\}$ present, for $s=1,2$.
\end{proposition}

\begin{corollary}\label{prop:WNBUic}
Consider a  $W$-model with heterogeneous servers, general service times and identical copies. 
The optimal policy may idle, but whenever a job is served,
it is stochastically better for the second-level policy to serve
according to ROS than FCFS. 
\end{corollary}
We note that for deterministic service times, there is no distinction between i.i.d. and identical copies. Hence, combining Figures~\ref{fig61}(a) and~\ref{fig4},  we observe that for a given policy $\pi$, the performance deteriorates as the variability of the service time increases. 

Finally, we note that having identical copies means that a copy in service always has a smaller remaining service time than a copy of this job that is not yet in service, as was the case for i.i.d. and NBU service times. Hence, we expect that Conjectures ~\ref{conj:ros-fcfs} and~\ref{conj:lrf-mrf-nbuiid} also hold when copies are identical and service times are general.

\begin{conjecture}\label{conj:ros-fcfs_identical}
Consider a redundancy system with a general topology, heterogeneous server capacities, general service times and identical copies, where idling is not permitted. We conjecture that for a given first-level policy, ROS outperforms (in some sense) FCFS as the second-level policy.
\end{conjecture}

\begin{conjecture}
\label{conj:lrf-mrf-identica}
Consider the nested redundancy model, heterogeneous server capacities, general service times and identical copies. We conjecture that for a given second-level scheduling policy $\Pi_2$,  it holds that 
$$\mathbb{E}(N^{LRF-\Pi_2})\leq \mathbb{E}(N^{MRF-\Pi_2}).
$$
\end{conjecture}

\subsection{Comparison between i.i.d. copies and identical copies}
\label{perf:cc}
In the present section we investigate how the correlation structure among the copies affects the performance of the system. For this section we define $N(t)$ as the total number of jobs in the system with i.i.d.\  copies, and $M(t)$ as the total number of jobs with identical copies. The proofs in this section are deferred to Appendix~A.

In \cite{Anton2019} the authors prove that the stability condition for the redundancy-$d$ (with $d>1$) model is larger under i.i.d. copies than under identical copies for FCFS. This is due to the fact that, when service times are exponential and copies are i.i.d., the departure rate of a subset of busy servers is given by the sum of all the service rates, whereas under identical copies it is given by the sum of the rates of servers giving service to different jobs. Hence, the departure rate under i.i.d.\ copies is at least as large as that under identical copies. In the following lemma, for a given policy $\Pi_1$-FCFS, we show a stronger result. For general service times and a general redundancy system, we show that the jobs with i.i.d.\ copies leave no later than with identical copies. 

\begin{proposition}\label{lemma1}
Consider a redundancy system with a general topology and heterogeneous server capacities, where the service times of the jobs are sampled from a general  distribution. Then, for any preemptive policy $\Pi_1$  we have that $\{ N^{\Pi_1-\textrm{FCFS}}(t)\}_{t\geq0}\leq_{st}\{ M^{\Pi_1-\textrm{FCFS}}(t)\}_{t\geq0}$.
\end{proposition}

We note that the previous proposition also holds when $\Pi_1$ is \emph{not} a \emph{strict} priority policy. In particular, it holds 
for $\Pi_0$, i.e.,
$\{ N^{\textrm{FCFS}}(t)\}_{t\geq0}\leq_{st}\{ M^{\textrm{FCFS}}(t)\}_{t\geq0}$.

We believe that the above result would hold as well for other policies than $\Pi_1$-FCFS and FCFS. However, we do not have a proof for this. We note that when comparing Figure~\ref{fig61} (b) and Figure~\ref{fig4} (c) (both for the $W$-model with a Weibull distribution), the mean number of jobs under i.i.d. copies is smaller than under identical copies. 


\section{Stability condition}
\label{Sec:stab}
In the present section we discuss the stability condition for our scheduling policies for exponential job sizes. In the particular cases where either FCFS or ROS is implemented and in the absence of a first-level policy, the stability condition has been characterized under various conditions, which we briefly summarize in Section~\ref{Sec:sub:stabfcfs}. These stability results are however no longer valid when a first-level policy is implemented. In Section~\ref{Sec:sub:stablrf} we discuss stability results for first-level policies LRF and MRF. We further note that Section~\ref{Sec:stab} is focussed on exponential service times. To the best of our knowledge, no explicit stability results have been obtained so far for ROS or FCFS when assuming general service times.

\subsection{Stability conditions for single-level FCFS and ROS} 
\label{Sec:sub:stabfcfs}
Under the FCFS service policy, the stability condition has been fully characterized in the case where jobs have exponentially distributed i.i.d.\ copies, \cite{Gardner16}. 

\begin{proposition}\label{Stab:Gard}\cite{Gardner16}
The redundancy heterogeneous server system with exponential job sizes and i.i.d. copies under FCFS, is stable if, for all $C\subseteq \mathcal C$,
\begin{equation}\label{eq:Stab:FCFS}
\lambda \sum_{c\in C} p_c < \sum_{s\in S(C)} \mu_s, 
\end{equation}
where $S(C)= \cup_{c\in C} \{s\in c\}$. 
\end{proposition}

This stability condition coincides with that of the \emph{cancel-on-start} redundancy system with exponential service times and FCFS. In addition, for exponential service times, Equation~(\ref{eq:Stab:FCFS}) is the maximum stability condition, i.e., there does not exist a policy (with or without redundancy) that makes the system stable if one of the inequalities of (\ref{eq:Stab:FCFS}) were not satisfied.

When copies are identical, the stability region under redundancy is reduced under FCFS. For example, for the redundancy-$d$ model with homogeneous servers, the maximum stability condition~(\ref{eq:Stab:FCFS}) simplifies to $\lambda<\mu K$ for i.i.d. copies, but for identical copies we have the stricter condition below ~\cite{Anton2021,Anton2019}.

\begin{proposition}\cite{Anton2021,Anton2019}
The redundancy-$d$ model under FCFS with exponential job sizes and identical copies is stable if $\lambda < \bar\ell\mu$ and unstable if $\lambda>\bar\ell\mu$, where $\bar\ell$ denotes the mean number of jobs in service in the associated saturated system. 
\end{proposition}

Under ROS, in \cite{Anton2019} the authors prove that for the redundancy-$d$ model under either i.i.d. copies or identical copies, the stability region is not reduced due to adding redundant copies.

\begin{proposition}
\label{th:ros_stab}
The redundancy-$d$ model with exponential job sizes and either i.i.d.\ or identical copies under ROS, is stable under condition~\eqref{eq:Stab:FCFS}, which reduces to  $\lambda<K\mu$.
\end{proposition}

\subsection{Stability condition under LRF and MRF}  
\label{Sec:sub:stablrf}
\subsubsection{I.I.D. copies}
We first assume the nested redundancy topology and that copies are i.i.d.. The stability condition under preemptive LRF-$\Pi_2$ with exponentially distributed service times is straightforward from Proposition~\ref{prop:lrf_mrf} and \cite{Gardner16}. In particular, the result shows that preemptive LRF-$\Pi_2$ is maximally stable. Below, we prove that the same holds true for non-preemptive LRF-$\Pi_2$.

\begin{proposition}\label{stab:lrf}
Consider a redundancy system with nested topology, where jobs are exponentially distributed with unit mean and have i.i.d. copies, and LRF-$\Pi_2$ is implemented. When the LRF policy is either preemptive or non-preemptive, the system is stable if for all $c\in\mathcal C$, 
\begin{equation}\label{eq:stab1} \lambda \sum_{\tilde c\subseteq c} p_{\tilde c} < \sum_{s\in c} \mu_s.
\end{equation}
The system is unstable if there exists $\hat c \in\mathcal C$ such that $ \lambda \sum_{\tilde c\subseteq \hat c} p_{\tilde c} > \sum_{s\in \hat c} \mu_s.$
\end{proposition}
We note that for a nested topology, the maximum stability condition of Equation~(\ref{eq:Stab:FCFS}) simplifies to (\ref{eq:stab1}).
Indeed, for a nested system, Proposition~\ref{Stab:Gard} reduces to verifying Condition~(\ref{eq:Stab:FCFS}) for every set~$C$ of nested classes. Note that a  class~$c$ is represented by a set of servers, hence one class can contain another class, which explains why $c\in C$ in~(\ref{eq:Stab:FCFS}) can be replaced by  $\tilde c \subseteq c$ in~(\ref{eq:stab1}).

\noindent\textit{Proof:} 
The stability result when the LRF policy is preemptive follows directly from Proposition~\ref{prop:lrf_mrf} and \cite{Gardner16}. From Proposition~\ref{prop:lrf_mrf}, the stability region under any two-level policy LRF-$\Pi_2$
must be at least as large as that under FCFS. For exponential service times, the latter is maximally stable \cite{Gardner16}, and hence, so is LRF-$\Pi_2$ with preemptive LRF. The proof when the LRF policy is non-preemptive can be found in Appendix~B.\qed
\bigskip

The assumption that the redundancy topology is nested is crucial for maximum stability to hold. To see this we refer to the following example where we consider a non-nested redundancy model and observe that LRF-FCFS and LRF-ROS are not maximally stable. 

\begin{example}\label{example3}\textbf{Non-nested model:}
Consider a $K=4$ server system with homogeneous capacities $\mu_s=1$, $s=1,\ldots,4$, and with the following redundancy topology: $\mathcal C=\{\{1,2\},\{1,3\},\{4\},\{3,4\},\{2,3,4\}\}$. We note that this particular topology is non-nested. {We chose $\vec p=\{0.15, 0.15, 0.3, 0.15, 0.25 \}$. Thus, the maximum achievable stability region is $\lambda<4$.}  In Figure~\ref{fig:stab:iid}, we plot the trajectory of the total number of jobs as a function of time for various loads and service policies. In the figure we observe that neither LRF-FCFS nor LRF-ROS are stable when $\lambda =3.8$, while FCFS and ROS do provide a stable system for $\lambda = 3.8$. We include plots with $\lambda=3.2$ and $\lambda = 4.1$  for comparison.

\begin{figure}[t!]
\centering
\begin{subfigure}{.49\textwidth}
	\centering
	\includegraphics[width=1.1\textwidth]{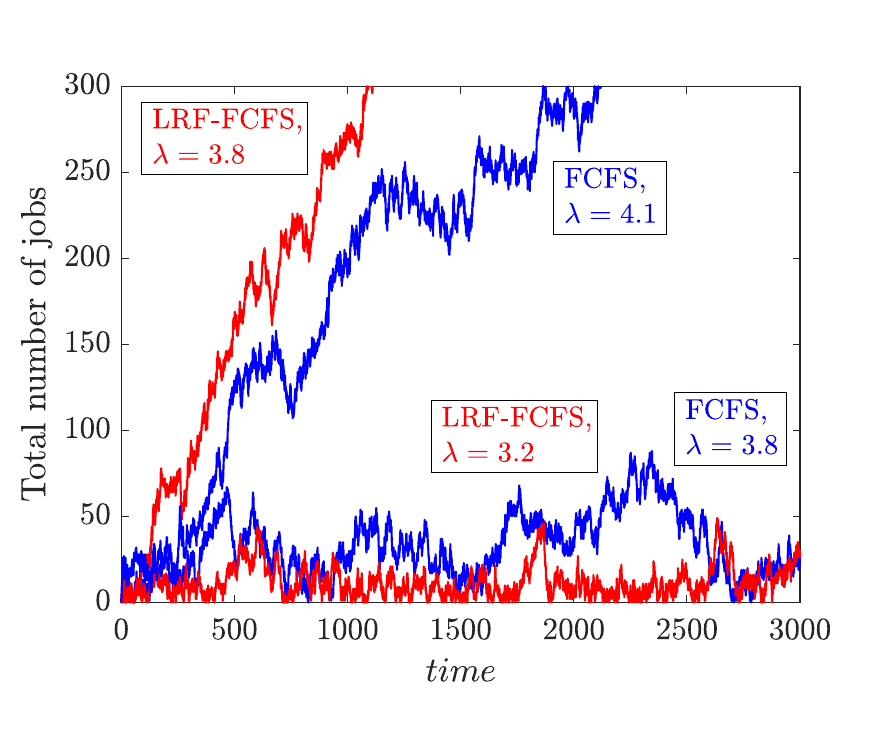}
	\subcaption{FCFS and LRF-FCFS}
\end{subfigure}
\begin{subfigure}{.49\textwidth}
	\centering
	\includegraphics[width=1.1\textwidth]{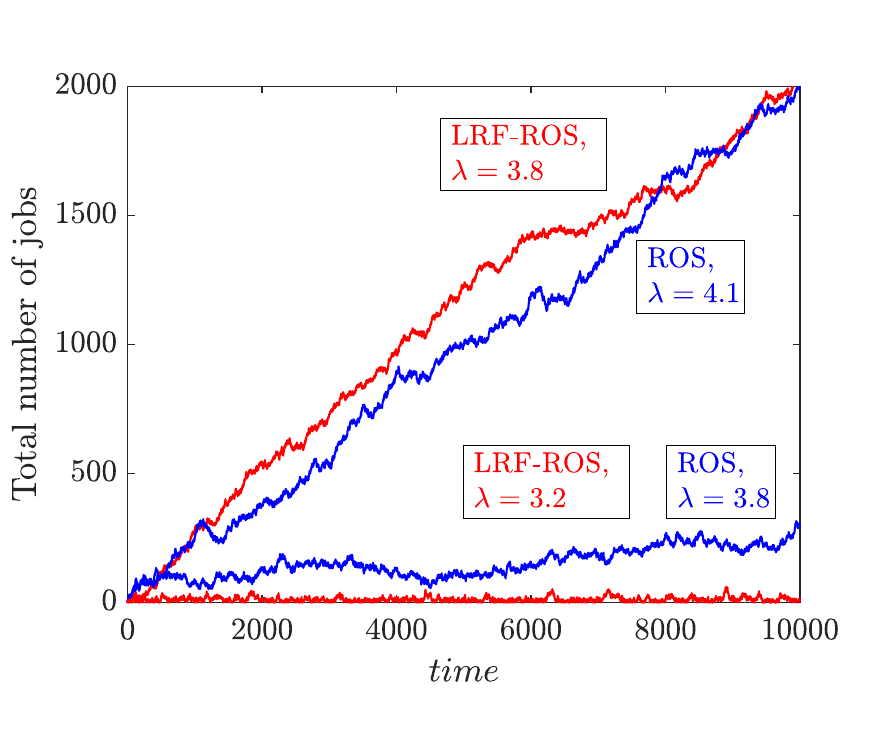}
	\subcaption{ROS and LRF-ROS}
\end{subfigure}
\caption{A non-nested redundancy topology with homogeneous server capacities. The trajectory of the total number of jobs for exponentially distributed service times and i.i.d. copies for scheduling policies $\Pi_1$-$\Pi_2$ with $\Pi_1$=LRF, $\empty$ and $\Pi_2$=FCFS, ROS for various arrival rates.}
\label{fig:stab:iid}	
\end{figure}
\end{example}

The assumption that $\Pi_1$=LRF is crucial in order for the maximum stability result to hold, {in addition to the assumption of nested structure}. To see this, we refer to Example~\ref{example21} where we will show that the nested $N$-model is not maximally stable under either MRF-FCFS or MRF-ROS. 

\begin{example}
\label{example21}\textbf{$N$-model:} We assume the $N$-model where servers have heterogeneous capacities $\vec\mu=(\mu_1,\mu_2)$. The maximum stability condition as given in (\ref{eq:Stab:FCFS}) when jobs have exponentially distributed service times simplifies to $\lambda p_{\{2\}}<\mu_2$ and $\lambda<\mu_1+\mu_2$. 

In the case of MRF-$\Pi_2$, with $\Pi_2$ non-idling, we have that class-$\{2\}$ jobs can only be served if there is no class-$\{1,2\}$ job present in the system. Let us denote by $\mu_{\{1,2\}}$ the mean departure rate of class-$\{1,2\}$ jobs in the system and by $\rho_{\{1,2\}} = \lambda (1-p_{\{2\}})/\mu_{\{1,2\}}$. Because copies are i.i.d., $\mu_{\{1,2\}}=\mu_1+\mu_2$, so the stability condition of class $\{1,2\}$ is $\lambda p_{\{1,2\}} <\mu_1+\mu_2$. Now class $\{2\}$ can only be served a $(1-\rho_{\{1,2\}})$ fraction of the time. Thus, the stability condition for class $\{2\}$ is given by $\lambda p_{\{2\}}<\mu_2(1-\rho_{1,2})$. 
This is a more strict condition than the maximum stability condition that only required $\lambda p_{\{2\}} <\mu_2$.%
\end{example}

In general, in order for a policy to be stable under the  maximum stability condition, the policy needs to correctly balance the jobs over the different heterogeneous servers so that in the long run, each class is in service an appropriate fraction of time. In general, this can be difficult. However, when the topology is nested, a job class only shares servers with classes that are subsumed within. As a result, when implementing LRF, each job class receives full capacity on its feasible servers when no higher priority jobs are present.

\subsubsection{Copies with general correlation structure}

{We now permit copies of a job to have arbitrary correlations, rather than the extreme cases of i.i.d. and identical (or perfect positive correlation). An example of a general correlation structure is the S\&X model of Gardner et al. \cite{Gardner17b} and Raaijmakers et al. \cite{Raaijmakers2018a} in which the runtime of each copy is decoupled into two components: the copy size $x$, which is the inherent size of the copy and is identical for all the copies of the same job, and the slowdown $s$, which is the slowdown experienced at the server where the copy is in service, and is independent across servers.}
More generally, we allow any joint distribution of job copies, but we do assume the marginal distribution for each copy is exponential.

Our first result is that for a nested topology, LRF-ROS is maximally stable. The proof is deferred to Appendix~B.

\begin{proposition}\label{stab:lrf-ros}
Consider a redundancy system with nested topology, where jobs are exponentially distributed with unit mean and copies follow some general correlation structure. Under LRF-ROS, with LRF either preemptive or non-preemptive, the system is stable if for all $c\in\mathcal C$, 
$$ \lambda \sum_{\tilde c\subseteq c} p_{\tilde c} < \sum_{s\in c} \mu_s.$$
The system is unstable if there exists $\hat c \in\mathcal C$ such that $ \lambda \sum_{\tilde c\subseteq \hat c} p_{\tilde c} > \sum_{s\in \hat c} \mu_s.$
\end{proposition}

For a nested redundancy topology with LRF, given the state of the system, the job class in service at each server is completely characterized. Furthermore, since the second-level policy is ROS, when the number of jobs in service is large, the probability that more than one copy of the same job is simultaneously in service is close to zero, regardless of the correlation structure among the copies. Hence, we can completely characterize the instantaneous departure rate of the system in the fluid limit. We note that for the {single-level} redundancy-oblivious policy ROS, the stability condition is unknown so far (with the exception of the redundancy-$d$ model).

For non-nested systems, or two-level policies other than LRF-ROS, we did not succeed in deriving the stability conditions. We do however show, in the examples below, some situations that might not be maximally stable. 

%
\begin{example}
\label{example2}\textbf{$N$-model:} We consider the $N$-model where servers have heterogeneous capacities $\vec\mu=(\mu_1,\mu_2)$ as in Example~\ref{example21}. We recall that the maximum stability condition when jobs have exponentially distributed service times is given by $\lambda p_{\{2\}}<\mu_2$ and $\lambda<\mu_1+\mu_2$. 

In the case of \textbf{MRF-$\Pi_2$} with \emph{identical} copies and $\Pi_2$ non-idling, the departure rate of class-$\{1,2\}$ is given by $\mu_{\{1,2\}}=\max\{\mu_1,\mu_2\}$. Because class $\{2\}$ can only be served a $(1-\rho_{\{1,2\}})$ fraction of the time, the stability condition is given by $\lambda p_{\{2\}} <\mu_{\{2\}}(1-\rho_{\{1,2\}})$, 
which is more strict than the maximum stability condition. 

In the case of \textbf{LRF-FCFS} with identical copies, when class $\{2\}$ and class $\{1,2\}$ are both present in the system, the total departure rate is given by $\mu_1+\mu_2$. 
Then, when class $\{2\}$ is not present (which happens $\rho_2=\lambda p_{\{2\}}/\mu_2$ fraction of the time), the total departure rate 
{for class $\{1,2\}$ jobs, call it $\tilde{\mu}_{\{1,2\}}$, will be at most $\max\{\mu_1,\mu_2$\}}.
Therefore, the stability condition is given by $\lambda p_{\{2\}}/\mu_2$, $\lambda<\mu_1+\mu_2$ and $\lambda p_{\{1,2\}} <\tilde \mu_{\{1,2\}}(1-\rho_2)$, where $\tilde\mu_{\{1,2\}}\leq\max\{\mu_1,\mu_2\}$, which is more strict than the maximum stability condition.  
\end{example}

We did not succeed in obtaining the stability conditions for second-level policies other than LRF-ROS. We do however have the following comparison result, which is a direct consequence of Proposition~\ref{prop:n_ident}.

\begin{corollary}\label{stab:mrf}
Consider a redundancy system with nested topology, where service times are exponentially distributed with unit mean and identical copies. The stability condition under preemptive MRF-ROS, is at least as large as that under preemptive MRF-FCFS.
\end{corollary}

We note that the above corollary does not give us exact values for stability conditions, since for both MRF-ROS and MRF-FCFS, the stability condition is unknown.

We now consider a numerical example (Example~\ref{example1}) of a non-nested system under LRF-ROS with identical copies and observe that it is not maximally stable.   
\begin{figure}[t!]
\centering
\begin{minipage}{.49\textwidth}
\centering
\includegraphics[width=1.1\textwidth]{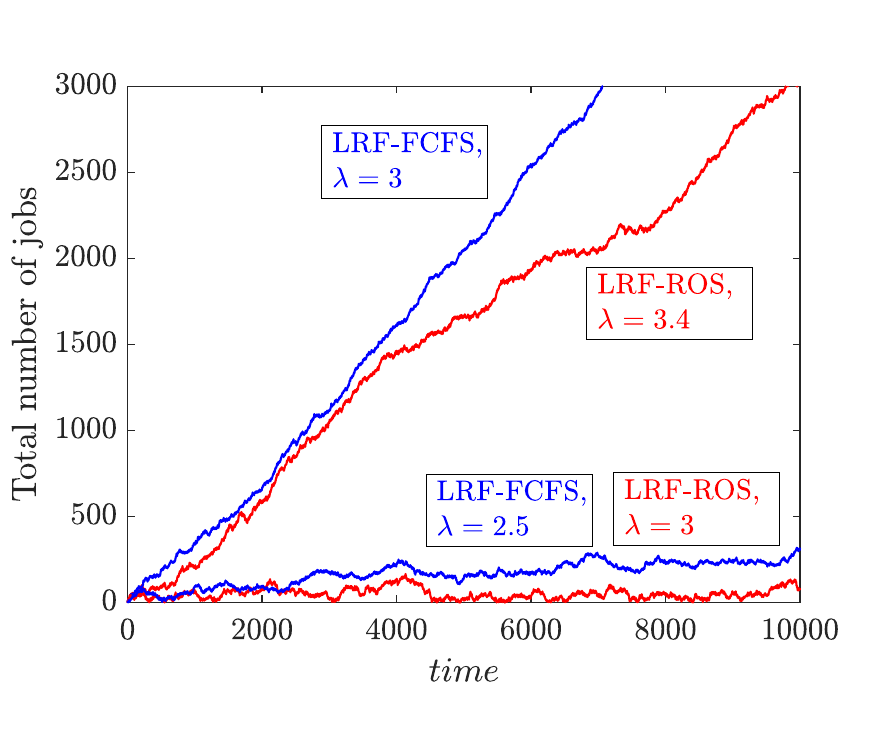}
\end{minipage}
\begin{minipage}{.49\textwidth}
\centering
\includegraphics[width=1.1\textwidth]{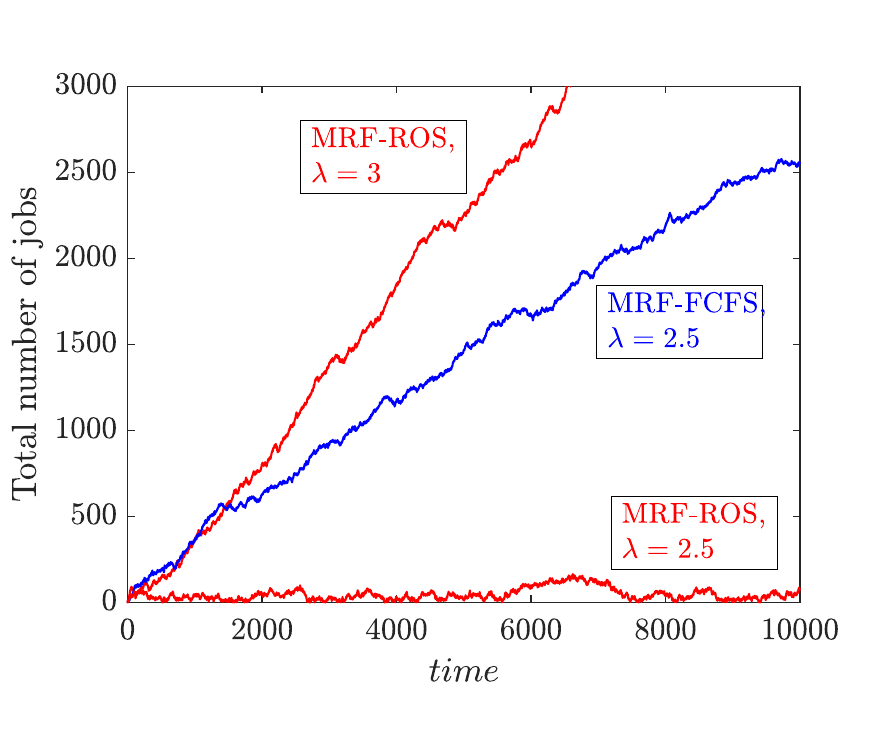}
\end{minipage}
\caption{Trajectory of the redundancy system of Example~\ref{example1} under different two-level policies and arrival rates. }
\label{fig_stab}	
\end{figure}

\begin{example}
\label{example1}\textbf{Non-nested model:}
We consider 4 homogeneous servers with unit capacities, identical copies, and jobs dispatch either 2 or 3 copies chosen uniformly at random. The maximum stability condition is $\lambda<4$ (\ref{eq:Stab:FCFS}). However, in Figure~\ref{fig_stab} we observe that already for $\lambda=3.4$ the system is not stable for any of the policies $\Pi_1$-$\Pi_2$, with $\Pi_1\in\{LRF,MRF\}$ and $\Pi_2\in\{FCFS,ROS\}$, including the LRF-ROS policy.
%
We also observe that the stability region under LRF-$\Pi_2$ is larger than that under MRF-$\Pi_2$. For LRF-ROS 
the stability region is at least $\lambda<3$, whereas for MRF-ROS, the system is unstable for $\lambda=3$.
We observe similar results for LRF-FCFS and MRF-FCFS at $\lambda=2.5$.
Finally, we observe  that the stability region under $\Pi_1$-ROS is larger than that under $\Pi_1$-FCFS, which 
is consistent with our result for $\Pi_1$ = MRF with nested topologies; see Corollary~\ref{stab:mrf}.  
\end{example}

\section{Redundancy-aware versus redundancy-oblivious\\ scheduling}
\label{aware}

In the previous sections, we obtained several (partial) optimality results for two-level policies. In this section we will {numerically} investigate whether these two-level policies are worth the trouble, that is, whether they can improve significantly the performance of the system compared to {single-level} redundancy-oblivious schedulers. We focus in this section on nested topologies, as most optimality results are for that context, and {such policies} can model server pools in data centres. 

\subsection{I.I.D. copies }

\begin{figure}[t!]
\centering
\begin{minipage}{.49\textwidth}
\centering
\includegraphics[width=1.1\textwidth]{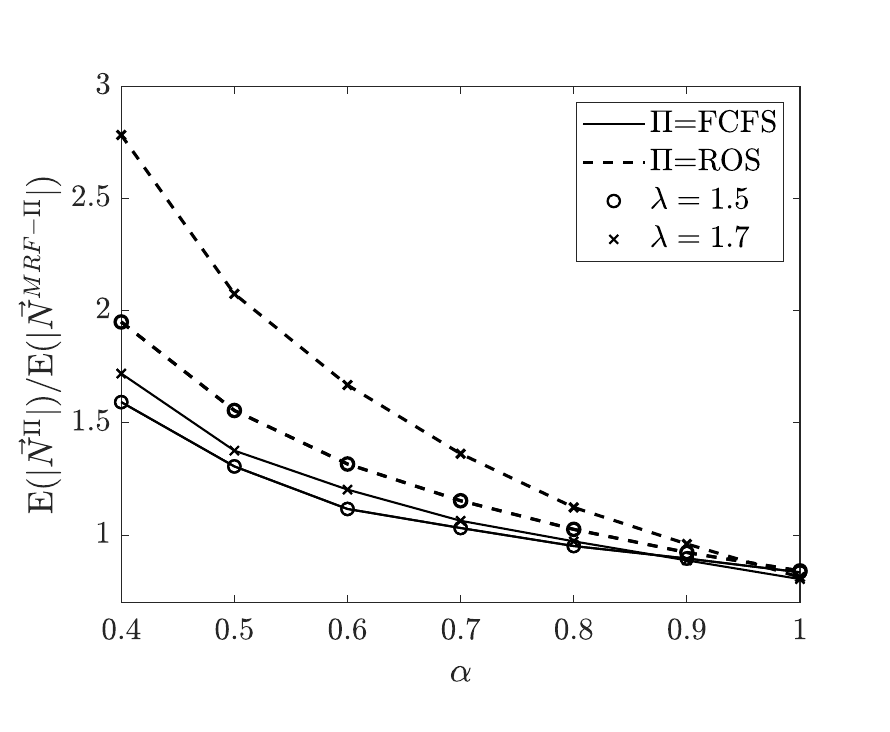}
\subcaption{$\alpha\leq 1$, NWU.}
\end{minipage}
\begin{minipage}{.49\textwidth}
\centering
\includegraphics[width=1.1\textwidth]{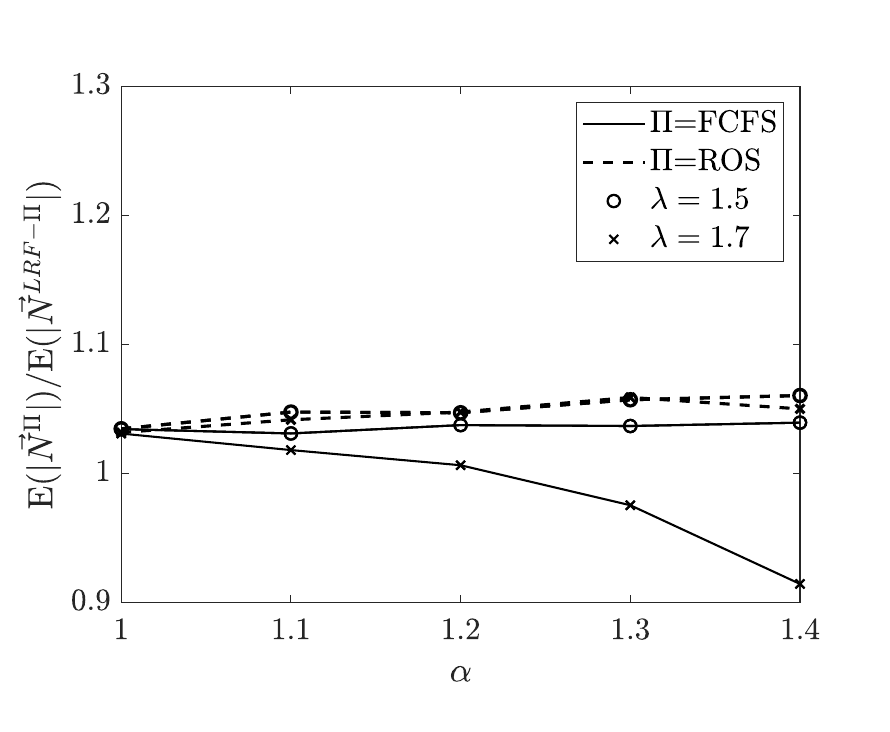}
\subcaption{$\alpha\geq 1$, NBU.}
\end{minipage}
\caption{The mean number of jobs for the W model with i.i.d. copies when
$\vec\mu=(1,1)$ and $\vec p=(1/6,1/6,2/3)$ with $X$ the Weibull distribution with parameter $\alpha$.}
\label{fig:Weibul_awarenes_iid}	
\end{figure}

For NWU service times, we observe from Figure~\ref{fig2} for the W model that the gap between $\Pi_1$-ROS and ROS, and the gap between $\Pi_1$-FCFS and FCFS, increases as the variability of the service time increases. That is, as $q$ goes to zero, or as the distribution $Y$ becomes more variable (i.e., moving from Figure~\ref{fig2}(a) to \ref{fig2}(b) to \ref{fig2}(c)). The redundancy-oblivious policies can be more than a factor 1.5 worse than the MRF redundancy-aware version. To further investigate this, in Figure~\ref{fig:Weibul_awarenes_iid} (a) we plot the performance as a function of the Weibull parameter $\alpha$, for $\alpha\leq 1$, that is for NWU distributions. As $\alpha$ decreases, the service time distribution becomes more variable. We compare the performance of redundancy-oblivious policies $\Pi_0$ against redundancy-aware policies  MRF-$\Pi_2$, {with $\Pi=\Pi_0=\Pi_2$}. We observe that redundancy-oblivious policies can have up to 2.5 times more jobs in the system than the MRF redundancy-aware policy for $\alpha =0.4$. 

On the other hand, for NBU service times, being redundancy aware matters less. For example, we can observe from Figure~\ref{fig61} that the gaps between ROS and LRF-ROS and between FCFS and LRF-FCFS are small, especially compared to the gap between ROS and FCFS. In Figure~\ref{fig:Weibul_awarenes_iid} (b) we plot the Weibull distribution for $\alpha \geq 1$ (hence NBU), and plot the ratio of the redundancy-oblivous policy $\Pi$ against the redundancy-aware policy LRF-$\Pi$. Again, the redundancy-oblivious policy performs very similarly to the LRF redundancy-aware version.

\subsection{Identical copies}
For identical copies, we observed in Section~\ref{subset:id} that LRF is an efficient first-level policy. We now first focus on Figure~\ref{fig4} and compare the performance of redundancy-aware policies LRF-$\Pi_2$ to that of redundancy-oblivious policies $\Pi_0$, for {$\Pi_0=\Pi_2=\Pi=$}FCFS, ROS. We observe from Figure~\ref{fig4} that the mean number of jobs under LRF-$\Pi_2$ is smaller than that under $\Pi_2$ for any value of $p_{\{1,2\}}$. We also note that this gap is more pronounced when the variability of the service time distribution is large. Note that the exponential distribution has squared coefficient of variation $C^2=1$ (Figure~\ref{fig4} (a) and (b)), the Weibull distribution with $\alpha=1.25$ has $C^2=0.64$ (Figure~\ref{fig4} (c) and (d)), and the degenerate hyperexponential distribution with $q=0.1$ has $C^2=19$ (Figure~\ref{fig4} (e) and (f)). Redundancy-oblivious policies can be more than a factor 1.5 worse than their  LRF redundancy-aware version.

\begin{figure}[t!]
\centering
\begin{minipage}{.49\textwidth}
\centering
\includegraphics[width=1.1\textwidth]{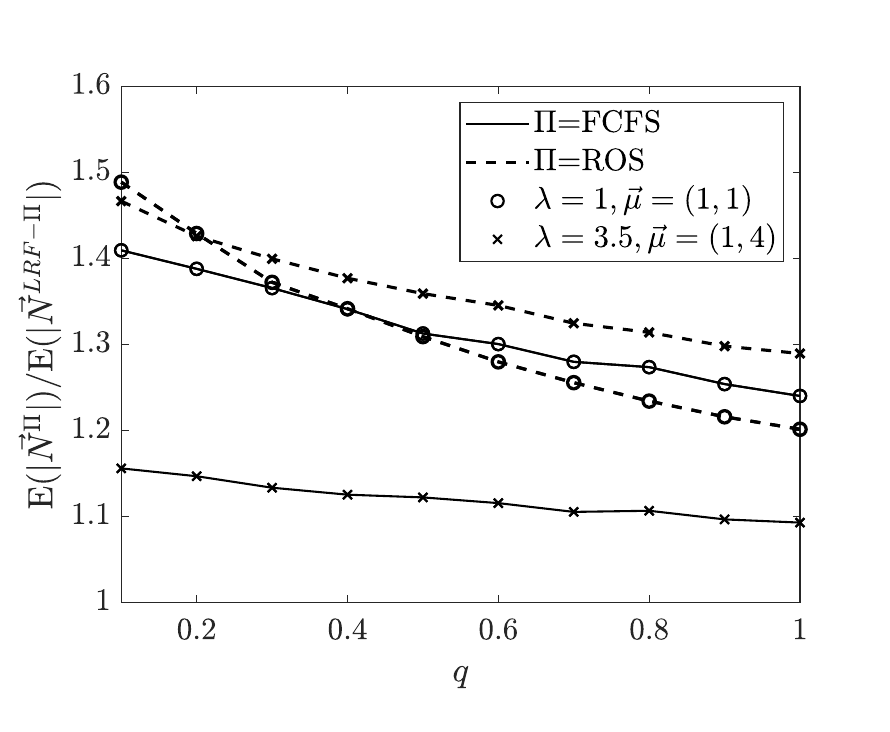}
\end{minipage}
\caption{The mean number of jobs for the $W$-model with identical copies when $\vec p=(0.2,0.2,0.6)$ and with $X$ mixture of exponential distribution ($q$) with respect to $q$.}
\label{fig:Weibul_awarenes}	
\end{figure}

In Figure~\ref{fig:Weibul_awarenes} we plot the ratio between the mean number of jobs under 
$\Pi_0$ and 
under LRF-$\Pi_2$, for {$\Pi_0=\Pi_2=\Pi=$} FCFS, ROS, when jobs have a mixture of exponential ($q$) distributed service times with respect to parameter $q$. We observe that the redundancy-oblivious policies are more than a factor 1.2 worse than the LRF redundancy-aware version. Moreover, we observe that this factor increases up to 1.5 as the variability of the service time distribution increases, that is, as $q$ goes to $0$. 

\section{Conclusions}
We have explored, for nested systems and cancel-on-complete redundancy, the performance impact of two-level policies based on the level of redundancy, and compared these policies with traditional single-level service disciplines such as FCFS. Our theoretical and numerical results indicate that FCFS is the best single-level policy and the best second-level policy when service times are NWU (so highly variable) and i.i.d. across copies. When variability is very high, it may be advantageous to use MRF as the first-level policy. When service time variability is low, or copies are identical, ROS is the best single-level policy and second-level policy, and LRF is the best first-level policy, though the first-level has less of an impact with low-variability services.

\section*{Acknowledgements}
We are grateful for the extensive feedback from an anonymous referee,
which improved the clarity of our exposition.

The research of E. Anton was carried out when she was at the Eindhoven University of Technology (TU/e), The Netherlands. Her research was partially supported by NWO through the Gravitation project NETWORKS under grant No 024.002.003, and has received funding from the European Union’s Horizon 2020 research and innovation program under the Marie Skłodowska-Curie grant agreement No 101034253. The research of I.M.\ Verloop was partially supported by the French ``Agence Nationale de la Recherche (ANR)'' through the project ANR-22-CE25-0013 (EPLER).

\bibliography{sn-bibliography}

\begin{thebibliography}{10}

\bibitem{Akgun2010}
Osman Akgun, Rhonda Righter, and Ronald Wolff.
\newblock Multiple server system with flexible arrivals.
\newblock {\em Applied Probability Trust}, 43: 985--1004, 2011.

\bibitem{Akgun2013}
Osman Akgun, Rhonda Righter, and Ronald Wolff.
\newblock Partial flexibility in routeing and scheduling.
\newblock {\em Advances in Applied Probability}, 45: 673--691, 2013.

\bibitem{Ananthanarayanan2012}
Ganesh Ananthanarayanan, Ali Ghodsi, Scott Shenker, and Ion Stoica.
\newblock Why let resources idle? aggressive cloning of jobs with dolly.
\newblock In {\em Proceedings of the 4th USENIX Conference on Hot Topics in
Cloud Computing}, HotCloud' 12, pp.~17, USA, 2012.

\bibitem{Ananthanarayanan13}
Ganesh Ananthanarayanan, Ali Ghodsi, Scott Shenker, and Ion Stoica.
\newblock Effective straggler mitigation: Attack of the clones.
\newblock In {\em NSDI}, 13: 185--198, 2013.

\bibitem{Anton2021}
Elene Anton, Urtzi Ayesta, Matthieu Jonckheere, and Ina~Maria Verloop.
\newblock A survey of stability results for redundancy models. In {\em Alexey B. Piunovskiy and Yi Zhang (eds.), Modern Trends in Controlled Stochastic Processes: Theory and Applications, V.III.} Springer US, 2021.

\bibitem{Anton2020}
Elene Anton, Urtzi Ayesta, Matthieu Jonckheere, and Ina~Maria Verloop.
\newblock Improving the performance of heterogeneous data centers through
redundancy.{\em Proceedings
of the ACM on Measurement and Analysis of Computing Systems (POMACS),
SIGMETRICS 2021}, 4(3): Article 48, pp. 29, 2020.

\bibitem{Anton2019}
Elene Anton, Urtzi Ayesta, Matthieu Jonckheere, and Ina~Maria Verloop.
\newblock On the stability of redundancy models.
\newblock {\em  Operations Research} 69(5): 1540--1565, 2021.

\bibitem{Bonald17a}
Thomas Bonald and C{\'e}line Comte.
\newblock Balanced fair resource sharing in computer clusters.
\newblock {\em Performance Evaluation}, 116: 70--83, 2017.

\bibitem{Bramson06}
Maury Bramson.
\newblock {\em Stability of Queueing Networks}.
\newblock Springer, 2008.

\bibitem{Dean13}
Jeffrey Dean and Luiz~Andr{\'e} Barroso.
\newblock The tail at scale.
\newblock {\em Communications of the ACM}, 56(2): 74--80, 2013.

\bibitem{Gardner2017}
Kristen Gardner, Mor Harchol-Balter, Esa Hyytia, and Rhonda Righter.
\newblock Scheduling for efficiency and fairness in systems with redundancy.
\newblock {\em Performance Evaluation}, 116: 1--25, 2017.

\bibitem{Gardner17b}
Kristen Gardner, Mor Harchol-Balter, Alan Scheller-Wolf, and Benny van Houdt.
\newblock A better model for job redundancy: Decoupling server slowdown and job
size.
\newblock {\em IEEE/ACM Transactions on Networking}, 25(6): 3353--3367, 2017.

\bibitem{Gardner17}
Kristen Gardner, Mor Harchol-Balter, Alan Scheller-Wolf, Mark Velednitsky, and
Samuel Zbarsky.
\newblock Redundancy-d: The power of d choices for redundancy.
\newblock {\em Operations Research}, 65: 1078--1094, 2017.

\bibitem{GHR19}
Kristen Gardner, Esa Hyyti\"{a}, and Rhonda Righter.
\newblock A little redundancy goes a long way: Convexity in redundancy systems.
\newblock {\em Performance Evaluation}, 131(4): 22--42, 2019.

\bibitem{Gardner2020}
Kristen Gardner and Rhonda Righter.
\newblock Product forms for fcfs queueing models with arbitrary server-job
compatibilities: An overview, {\em Queueing Systems,} 96(1-2): 3--51, 2020.

\bibitem{Gardner16}
Kristen Gardner, Samuel Zbarsky, Sherwin Doroudi, Mor Harchol-Balter, Esa
Hyyti\"{a}, and Alan Scheller-Wolf.
\newblock Queueing with redundant requests: exact analysis.
\newblock {\em Queueing Systems}, 83(3-4):227--259, 2016.

\bibitem{HB13}
Mor Harchol-Balter.
\newblock {\em Performance Modeling and Design of Computer Systems: Queueing
Theory in Action}.
\newblock Cambridge University Press, 2013.

\bibitem{HvH18}
Tim Hellemans, and Benny  van Houdt:
\newblock Analysis of redundancy(d) with identical replicas.
\newblock {ACM Sigmetrics Performance Evaluation Review} \textbf{46(3)}, 74--79 (2018)


\bibitem{Joshi15}
Gauri Joshi, Emina Soljanin, and Gregory Wornell.
\newblock Queues with redundancy: Latency-cost analysis.
\newblock {\em ACM SIGMETRICS Performance Evaluation Review}, 43(2): 54--56,
2015.

\bibitem{Kim2009}
Yusik Kim, Rhonda Righter, and Ronald Wolff.
\newblock Job replication on multiserver systems.
\newblock {\em Advances in Applied Probability}, 41: 546--575, 2009.

\bibitem{KR08}
Ger Koole and Rhonda Righter.
\newblock Resource allocation in grid computing.
\newblock {\em Journal of Scheduling}, 11: 163--173, 2008.

\bibitem{Lee17a}
Kangwook Lee, Nihar~B. Shah, Longbo Huang, and Kannan Ramchandran.
\newblock The mds queue: Analysing the latency performance of erasure codes.
\newblock {\em IEEE Transactions on Information Theory}, 63(5):2822--2842,
2017.

\bibitem{Mendelson2020}
Gal Mendelson:
\newblock A lower bound on the stability region of redundancy-d with  FIFO
service discipline.
\newblock {Oper. Res. Lett.} \textbf{49(1)}, 113--120 (2021)

\bibitem{Nageswaran}
Leela Nageswaran and Alan Scheller-Wolf.
\newblock Queues with redundancy: Is waiting in multiple lines fair?, {\em Manufacturing \& Service Operations Management}, 2021.

\bibitem{Raaijmakers2018a}
Youri Raaijmakers, Sem Borst, and Onno Boxma. \newblock{Delta probing policies for redundancy}. {\em Performance Evaluation,} 127-128, 2028.

\bibitem{Raaijmakers2018}
Youri Raaijmakers, Sem Borst, and Onno Boxma. \newblock Redundancy scheduling with scaled Bernoulli service requirements, {\em Queueing Systems Volume} 93:1-2, 2019.


\bibitem{Raaijmakers2019}
Youri Raaijmakers, Sem Borst, and Onno Boxma:
\newblock Stability of redundancy systems with processor sharing.
\newblock In: {Proc. of the 13th EAI Intern. Conf. on
Performance Evaluation Methodologies and Tools}, Valuetools 20, pp. 120--127 (2020)

\bibitem{Raaijmakers2020}
Youri Raaijmakers and Sem Borst.
\newblock Achievable stability in redundancy systems, {\em Proc. ACM Meas. Anal. Comput. Syst.} 4(3): Article 46, 2020.

\bibitem{Robert03}
Philippe Robert.
\newblock {\em Stochastic Networks and Queues}.
\newblock Springer-Verlag, 2003.

\bibitem{Shah16}
Nihar~B. Shah, Kangwook Lee, and Kannan Ramchandran.
\newblock When do redundant requests reduce latency?
\newblock {\em IEEE Transactions on Communications}, 64(2): 715--722, 2016.

\bibitem{Vulimiri13}
Ashish Vulimiri, Philip~Brighten Godfrey, Radhika Mittal, Justine Sherry,
Sylvia Ratnasamy, and Scott Shenker.
\newblock Low latency via redundancy.
\newblock In {\em Proceedings of the ninth ACM conference on {E}merging networking
experiments and technologies}, 283--294, 2013.

\end{thebibliography}




\section*{Appendix}
\label{Appendix}

\subsection*{A: Proofs of Section~\ref{perf}}

\noindent\textbf{Proof of Proposition~\ref{lem:NWUiid}:}
We prove that when there are two jobs of the same class to be served, it is always better to serve a copy of the job that already has a copy in service in another server than to serve a copy of the job that has no other copies in service. This is a stronger claim than that of the proposition. 
The argument is similar to that of \cite{KR08}, but we sketch the proof here for completeness. 

We assume that at some time $t$ under policy $\pi$ a server starts to serve a copy of a job that has no other copies in service, say this copy is from job 2 and in server 1. Additionally, there is a job at server 1 of the same class as job 2 that has copies being served by another server(s), say that this is job 1.  Let ${\cal A}$ be the set of servers that job 1 has already received some service on and let ${\cal B}$ be the set of servers that can serve job 1 but have not yet started to serve job 1. 
{Note that servers in ${\cal A}$ will continue to give priority to job 1 over job 2 by our assumption on $\Pi_2$ being non-preemptive.}

\begin{figure}
\centering
\includegraphics[scale=0.4]{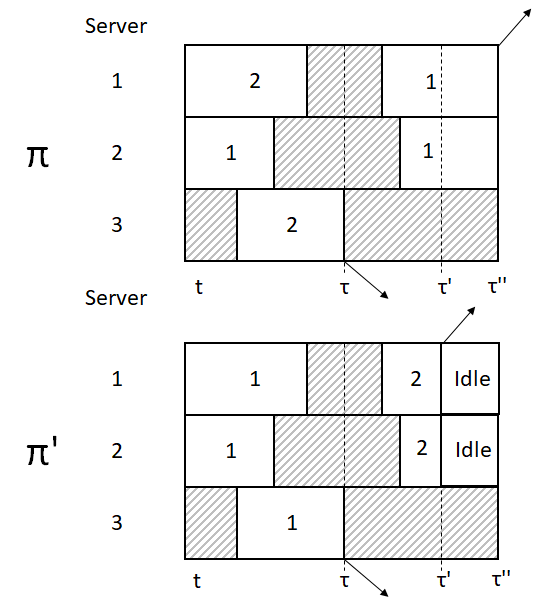}
\caption{Example for the proof of Proposition~\ref{lem:NWUiid}.}
\label{fig:Prop3_1}
\end{figure}

We let $\pi ^{\prime }$ serve job 1 on server 1 at time $t$ and thereafter always
serve job 1 whenever $\pi $ serves job 2 or job 1 on any server in ${\cal B}$, until either job 1 or job 2 completes service under $\pi $; {let us call that} time $\tau $. 
Otherwise we let $\pi ^{\prime }$ agree with $\pi $ until time $\tau $
(including when $\pi $ serves job 1 on servers in ${\cal A}$). We couple the service times of the copies of jobs 1 and 2 on servers in $\mathcal{B}$ under the two policies, and let all other service times and arrival times be the same under both policies, {so that job 1 completes under $\pi ^{\prime}$ at time $\tau$}. 
At time $\tau$, under policy $\pi ^{\prime }$, job 1 has completed service and job 2 has not yet received service at any of its compatible servers. Under policy $\pi $, either job 1 or job 2 has completed service. Let us denote by $a$ the job that did not complete service under $\pi$, $a$ is either job~1 or job~2. We note that job~$a$ received service partially at some servers under policy $\pi$, say servers in the set $\cal J$. We couple the (new) service times of job 2
under $\pi ^{\prime }$ on servers $\cal J$, denoted by $X_{2j}^{\prime }$ for $j=1,\ldots,|\cal J|$, with the  
remaining service times of job $a$, denoted by $X_{aj}$ for $j=1,\ldots,|\cal J|$, so that, $X_{2j}^{\prime
}\leq X_{aj}$ with probability 1 for all $j=1,\ldots,|\cal J|$. We can do this from the NWU assumption. We let $\pi^{\prime }$ serve  a copy of job 2 whenever $\pi $ serves a copy of job $a$ from time $\tau $ on, until job 2 completes under $\pi ^{\prime }$, 
{at some time $\tau ^{\prime} > \tau$}. Thereafter, we let the servers that are serving a copy of job $a$ under $\pi$ idle under $\pi'$, and otherwise let $\pi'$ agree with $\pi $,
{until job $a$ completes under $\pi$, at some time $\tau ^ {\prime \prime} \geq \tau ^ {\prime}$}. 
Then $\{\vec{N}^{\prime }(s)\}_{s\geq 0}\leq \{\vec{N}(s))\}_{s\geq 0}$ with probability 1.
{See Figure~\ref{fig:Prop3_1} for an example sample path Gantt chart with three servers that can all serve job 1, and where $a=1$. In the figure, shaded areas indicate times in which servers are serving jobs with higher priorities than jobs 1 or 2  and an arrow indicates a copy (and job) completion from the indicated server.}

Repeating the argument gives that FCFS, possibly with idling, is optimal. Then, we are left to verify that the non-idling FCFS policy is better than with idling FCFS.

Assume that at time $t$, $\pi$ idles a server, {say server 1,} for some time $\tau$ when it has copies to serve.
At time $t$, let $\pi'$ serve the copy of job 1 on server 1,
{and, coupling all arrival times and service times}, let it continue to serve job 1 on server 1 until 
time $t+\sigma$ where $\sigma=\min\{\tau,s, r-t,a-t\}$, $s$ is the remaining service time of the copy of job 1 on server 1, $r$ is the earliest completion time of all other copies of job 1 running on other servers and $a$ is the arrival time of a higher priority job than job 1. We let $\pi'$ agree with $\pi$ for all other decisions between times $t$ and $t+\sigma$. 
Four different events can occur at time $\sigma$:

\begin{itemize}
\item $\sigma=\tau$.  We let $\pi ^{\prime }$ idle server 1
after time $\sigma$ whenever $\pi $ serves job 1 on server 1, until
it has received service for $\tau$ units of time in that server  under $\pi$, at some time $\sigma' > \sigma$. 
Let $\pi ^{\prime }$ otherwise agree with $\pi $ for all
time after $\sigma$ and all servers. Then the systems under the two
policies will be in the same states at time $\sigma'$, and $\{N^{\prime
}(s)\}_{s\geq 0}=\{N(s)\}_{s\geq 0}$ with probability 1.

\item $\sigma=s$. We let $\pi ^{\prime }$ idle the servers serving job 1 after
time $\sigma$ whenever $\pi $ serves job 1 on those servers, until job 1 departs under $\pi $  at time $\sigma'$, say, and let $\pi'$ otherwise agree with $\pi $. We let 	$\pi ^{\prime }$ agree with $\pi $ thereafter.
At time $\sigma'$, the two systems will be in the same
state, but job 1 will have departed earlier under $\pi ^{\prime }$, so $\{N^{\prime }(s)\}_{s\geq 0}\leq \{N(s)\}_{s\geq 0}$ with probability 1.

\item $\sigma=r-t$. Then the job departs at time $\sigma$ in both systems and they will be in the same state. Letting $\pi ^{\prime }$ agree with $\pi $ after time $\sigma$, $\{N^{\prime }(s)\}_{s\geq0}=\{N(s)\}_{s\geq 0}$ with probability 1.

\item $\sigma=a-t$. At time $a$, job 1 is preempted in $\pi'$. Then, we let $\pi'$ idle server 1 after time $a$ whenever $\pi$ serves job 1 on server 1, until it has received service for $a-t$ units of time in that server (like job 1 received under policy $\pi'$). Let $\pi'$ otherwise agree with $\pi$ for all time after $a$. Then the systems under the two policies will be in the same states at time $a$, and $\{N^{\prime }(s)\}_{s\geq0}=\{N(s)\}_{s\geq 0}$ with probability 1.
\end{itemize}
\qed
\bigskip


\noindent\textbf{Proof of Proposition~\ref{prop:Wiid}:}
For ease of notation we remove the second-level policy from the superscript, which is FCFS under both systems. 
We note that for the random variable $X$, $\mathbb E(X) = \mathbb E(Y)$ and $\mathbb E(X^2)=\mathbb E(Y^2)/q$. Let $Z:=\min_s\{Y_i/\mu_i\}$. 

For a two-class single-server queue where class~1 has preemptive priority over class~2, the mean number of class-$k$ jobs is given by~\cite{HB13} {Chapter 32.} 
\begin{equation}\label{eq:formula}
\mathbb E(N(k)) = \lambda p_k\left[\frac{\mathbb E(X_k)}{1-\sum_{i=1}^{k-1} \rho_i} + \frac{\sum_{i=1}^k \rho_i \frac{\mathbb E(X_i^2)}{2\mathbb E(X_i)}}{(1-\sum_{i=1}^{k-1}\rho_i)(1-\sum_{i=1}^k\rho_i)}\right], \  k=1,2,
\end{equation}
where $X_k$ is the service time distribution of class-$k$ jobs, $\rho_k = \lambda_k\mathbb E(X_k)$, and $\lambda_k=p_k\lambda$ is the arrival rate of class-$k$ jobs, $k=1,2$. 

From the above result, we can compute the mean number of jobs under MRF-FCFS. We note that class-$S$ jobs have preemptive priority over all other classes. Let us denote by $U=\min_s\{X_s/\mu_s\}$ the service time of class-$S$ jobs under the  MRF  policy. Thus, $U=0$ with probability $1-q^K$ and $U=\frac{Z}{q}$ with probability $q^K$. The latter implies that  $\mathbb E(U) = q^{K-1} \mathbb E(Z)$ and
$\mathbb E(U^2) = q^{K-2} \mathbb E(Z^2)$.
By applying Equation~(\ref{eq:formula}) we obtain
\begin{eqnarray}\label{eq:MRF_S}
\mathbb{E}(N^{\rm{MRF}}_{S}) & = &\lambda p_{\{S\}}\left( \mathbb E(U) + \frac{\lambda p_{\{S\}}\mathbb E(U^2)}{2(1-\lambda p_{\{S\}}\mathbb E(U))} \right)\nonumber \\
& = &\lambda p_{\{S\}}\left( q^{K-1}\mathbb E(Z) + \frac{\lambda p_{\{S\}}q^{K-2}\mathbb E(Z^2)}{2(1-\lambda p_{\{S\}}q^{K-1}\mathbb E(Z))} \right)  .
\end{eqnarray}
Class-$\{s\}$ jobs are served if there are no class-$S$ jobs present. Let us denote by $W_s$ a generic service time of class-$\{s\}$ job, so $W_s = Y/\mu_s$, $\mathbb E(W_s) = \mathbb E(Y)/\mu_s$ and $\mathbb E(W^2) = \mathbb E(Y^2)/(q\mu_s^2)$. By applying Equation~(\ref{eq:formula}), we obtain
\bigskip

\noindent $\mathbb E(N_{\{s\}}^{MRF})$
\begin{eqnarray*}
&=&\frac{\lambda p_{\{s\}}}{1-\lambda p_{S}\mathbb E(U)}\left( \mathbb E(W_{s})+%
\frac{\lambda p_{S}\mathbb E(U^{2})+\lambda p_{\{s\}}\mathbb E(W_{s}^{2})}{2(1-\lambda
p_{S}\mathbb E(U)-\lambda p_{\{s\}}\mathbb E(W_{s}))}\right)  \\
&=&\frac{\lambda p_{\{s\}}}{1-\lambda p_{S}q^{K-1}\mathbb E(Z)}\left( \mathbb E(Y)/\mu _{s}+%
\frac{\lambda p_{S}q^{K-2}\mathbb E(Z^{2})+\lambda p_{\{s\}}\frac{1}{q}\mathbb E(Y^{2})/\mu _{s}^{2}}{%
2(1-\lambda p_{S}q^{K-1}\mathbb E(Z)-\lambda p_{\{s\}}\mathbb E(Y)/\mu _{s})}\right) 
\end{eqnarray*}
for $s=1,\ldots,K$. We therefore obtain, 
\begin{eqnarray}\mathbb{E}(N^{\rm{MRF}})  && =   \sum_{s\in S}\mathbb{E}(N^{\rm{MRF}}_{\{s\}}) +  \mathbb{E}(N^{\rm{MRF}}_{S}) \nonumber\\
&&= \frac{1}{q}\sum_{s=1}^K\frac{\lambda^2p_{\{s\}}^2\mathbb E(Y^2)/\mu_s^2}{2(1-\lambda p_{\{s\}}\mathbb E(Y)/\mu_s)} +o(1/q) 
\end{eqnarray}

Under LRF, class~$s$ receives preemptive priority, so that their mean queue length is given by 
\begin{eqnarray*}
\mathbb{E}(N^{\rm{LRF}}_{\{s\}}) & =& \lambda p_{\{s\}} \left( \mathbb E(W_s)+\frac{\lambda p_{\{s\}} \mathbb (W_s^2)}{2(1-\lambda p_{\{s\}}\mathbb E(W_s))} \right) \\
&= & \lambda p_{\{s\}}\left(\frac{\mathbb E(Y)}{\mu_s} +\frac{1}{q}\frac{\lambda p_{\{s\}} \mathbb E(Y^2)/\mu_s^2}{2(1-\lambda p_{\{s\}}\mathbb E(Y)/\mu_s)}\right).
\end{eqnarray*} 
Hence,  $$\sum_{s\in S}\mathbb{E}(N^{\rm{LRF}}_{\{s\}}) =   \frac{1}{q}\sum_{s=1}^K\frac{\lambda^2p_{\{s\}}^2\mathbb E(Y^2)/\mu_s^2}{2(1-\lambda p_{\{s\}}\mathbb E(Y)/\mu_s)} +o(1/q),$$ where we note that the term multiplied by $1/q$ coincides with that of $\mathbb{E}(N^{\rm{MRF}})$. In order to conclude the proof, it remains to prove that $\mathbb{E}(N^{\rm{LRF}}_{S})=C /q+o(1/q),$ with $C>0$.  

Class-$S$ jobs see a single-server queue where the capacity depends on the  number of classes of dedicated traffic that are present. For example, when a class-$S$ job enters service and there are $n$ classes of dedicated traffic present, it can be served in $K-n$ servers. 
This dependence on the dedicated traffic makes it  infeasible to obtain a closed-form expression for the mean number of class-$S$ jobs under LRF. We therefore consider the following lower bound on $N^{LRF}_S(t)$. Consider a single-server system with capacity~$1$ and only class-$S$ jobs. The server is on vacation whenever in the original LRF system all servers are working on dedicated traffic. The service time of a class-$S$ job is distributed according to $Z/q$ with probability $q^K$, and zero otherwise. This single-server system is a lower bound on $N^{LRF}_S(t)$, since (i) whenever class~$S$ is served in the original system, it is also served in the lower-bound system, and (ii) the generic service time of a job in the lower bound system is less than or equal to the generic service time in the original system in the server in which a copy of this job finishes service.

In the lower-bound system, we bound the mean sojourn time for this single-server queue with server interruptions following the  ``tagged job" technique as in \cite{HB13} ({Chapter 23}). That is, we imagine that a job, called the tagged job, enters the system and consider all the work that must be completed before it can leave. This is larger than the time it takes until class-$S$ can be served. This time is either zero (in case the tagged job arrived outside a vacation) or is distributed as $V$, where the random variable $V$ denotes the time left of the vacation. Hence, $\mathbb E(N_S^{LB})\geq \lambda p_S  \mathbb P(\textrm{the tagged job finds the server interrupted}) \mathbb E(V)$.

We first compute the probability that the tagged job finds the server interrupted. The LB system is interrupted as long as all servers are busy in the original system. For the original system, the long-run proportion of time server~$s$ is busy serving class-$\{s\}$ jobs is $\lambda p_{\{s\}} \mathbb E(W_s) = \lambda p_{\{s\}} \mathbb E(Y)/\mu_s$. Because classes~$1,\ldots,K$ are served independently, the long-run proportion of time that all servers are busy serving these classes is $\tilde p:= \Pi_{s\in S}\lambda p_{\{s\}} \mathbb E(W_s) = \Pi_{s\in S}\lambda p_{\{s\}} \mathbb E(Y)/\mu_s$.

The interruption of the LB system is completed as soon as in the original system there is a server that completes all its class-$\{s\}$ jobs. The time that each server needs to complete its work of class $\{s\}$, $T_s/\mu_s$ is lower bounded by the time that the server needs to serve the job in service. Because the service times of jobs with a strictly positive service time are NWU, $T_s$ is stochastically larger than  $Y_s/q$. Hence,
$$\mathbb E(N_S^{LRF})\geq   \lambda p_S  \tilde p \mathbb E(V) \geq \lambda p_S  \tilde p \mathbb E(\min_s\{T_s/\mu_s\}) \geq \frac{\lambda p_S \tilde p}{q}\mathbb E(\min_s\{Y_s/\mu_s\}),$$ that is, $\mathbb E(N_S^{LRF})$ is lower bounded by some strictly positive constant divided by $q$. 
\qed
\bigskip


\noindent\textbf{Proof of Proposition~\ref{prop:iid_NBU_dedicated_nonidling}:} 
Suppose a two-level policy $\pi $ idles server~$s$ at time $t$ when there is a job of class $\{s\}$, call it job 1, with highest priority under $\pi $. Note that, without loss of generality, we can label the next job served under $\pi$ job 1 because the second-level policy is nonpreemptive. That is, either job 1 will be the only class $\{s\}$ job that might have already received some service at time $t$, so will remain the highest priority class-$\{s\}$ job, or, because the class can only run on a single server, jobs that have not received some service time can be served in any order.
Let $\tau >t$ be the time that $\pi $ stops idling, either because it serves a
higher priority job that has arrived between times $t$ and $\tau $, or because it starts serving a class-$\{s\}$ job. Let $\pi ^{\prime }$ serve job 1
from time $t$ until it either completes or time $\tau $, whichever comes first, and let it agree with $\pi $ for the other servers until time $\tau$. Couple all arrival times and all service times under the two policies. 
If job 1
completes before time $\tau $ under $\pi ^{\prime }\,$, then letting $\pi ^{\prime }$ idle server $s$ whenever $\pi $ serves job 1 on server $s$, and
otherwise letting $\pi ^{\prime }$ agree with $\pi $, we have $\{N^{\prime}(s)\}_{s\geq 0}\leq _{st}\{N(s)\}_{s\geq 0}$. If job 1 does not complete by
time $\tau $, then we let $\pi ^{\prime }$ serve job 1 on server $s$
whenever $\pi $ does until job 1 completes under $\pi ^{\prime }$, at time $\tau ^{\prime }$, which will be before it completes under $\pi $. The rest
of the argument is the same as the previous case, but starting from time $\tau ^{\prime }$ instead of $\tau $.
\qed
\bigskip

\noindent\textbf{Proof of Proposition~\ref{prop:iid_NBU_2_FCFS}:} 
\begin{figure}
\centering
\begin{subfigure}[h]{0.45\textwidth}
\includegraphics[scale=0.4]{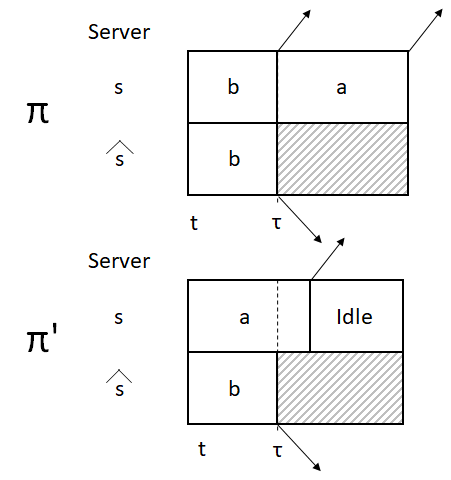} 
\subcaption{Case (i)}
\end{subfigure}
\begin{subfigure}[h]{0.45\textwidth}
\includegraphics[scale=0.4]{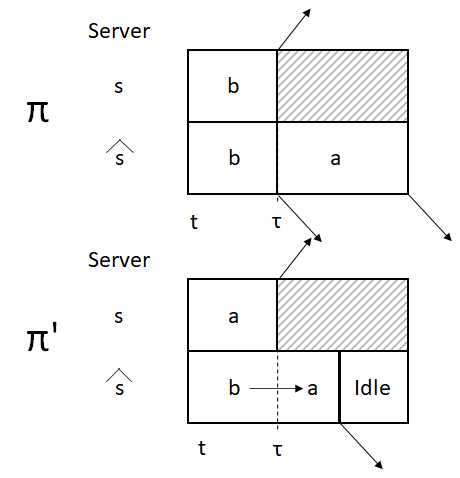}
\subcaption{Case (ii)}
\label{fig:Prop_cas2}
\end{subfigure}
\caption{Description of the two cases for the proof of Proposition~\ref{prop:iid_NBU_2_FCFS}}
\label{fig:Prop_cas}
\end{figure}

We assume that at time $t$ under policy $\pi $, class $\{1,2\}$ has highest priority at server $s$, and that $\pi $ starts serving a job, call it job $b$, on server $s$, where a copy of job $b$ has already received some service on the other server $\hat{s}$, and there is another class-$\{1,2\}$ job, call it job $a$, that has not yet received any service. Let $\pi ^{\prime }$ serve job $a$ instead of job $b$ on server $s$ whenever $\pi$ serves job $b$ on server $s$ and otherwise agree with $\pi $ until job $b$ completes under $\pi $, at time $\tau$ say. We couple all arrival and service times, including the service time on server $s$ starting at time $t$. If job $b$ completes on server $\hat{s}$ at time $\tau $ under $\pi $, {case (i) in Figure~\ref{fig:Prop_cas}~(a)}, then it also completes on server $\hat{s}$ under $\pi ^{\prime }$.  Due to the NBU assumption, we can couple the remaining service time  of job $a$ on server $s$ under policy $\pi'$ such that it is with probability 1 smaller than the service time of job $a$ on server s under policy $\pi$. Now, letting $\pi ^{\prime }$ agree with $\pi $ except for possibly idling when $\pi $ is serving job $a$ but job $a$ has completed under $\pi ^{\prime }$, we again have $\{N^{\prime }(s)\}_{s\geq 0}\leq_{st}\{N(s)\}_{s\geq 0}$. If job $b$ completes under $\pi $ on server $s$ at time $\tau $, 
{case (ii) in  Figure~\ref{fig:Prop_cas}~(b)}, then job $a$ completes under $\pi ^{\prime }$ on server $s$. Let us relabel job $b$ under $\pi ^{\prime }$ as job $a$. Then the systems under $\pi $ and $\pi ^{\prime }$ are in the same state, except that job $a$ has received some service under $\pi ^{\prime }$ and not under $\pi $. Arguing as before, $\{N^{\prime }(s)\}_{s\geq 0}\leq _{st}\{N(s)\}_{s\geq 0}$.

\qed
\bigskip

\begin{figure}
\centering
\includegraphics[scale=0.5]{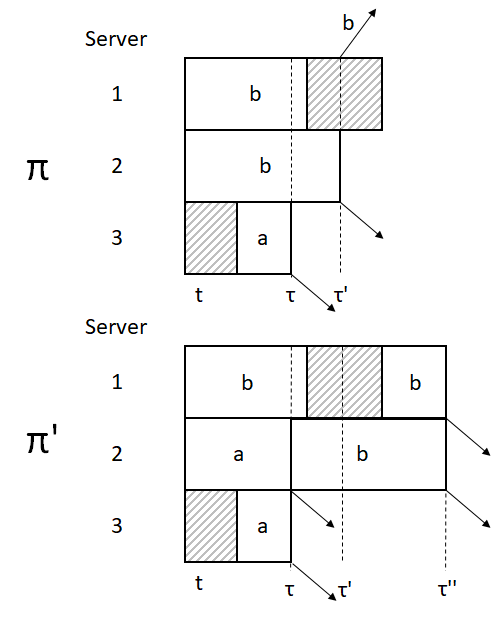}
\caption{A 3-server system Gantt chart where job $a$ leaves both systems $\pi$ and $\pi'$, but job $b$ does not.}
\label{fig:Prop_cas3}
\end{figure}

We now explain why Proposition~\ref{prop:iid_NBU_2_FCFS} cannot be generalized to more than two servers: At time $t$, there is a job in server 1 that has received some service, call this  job~$b$. We assume that under policy $\pi$, server 2 serves a fresh copy of job~$b$, and that under policy $\pi'$,  server 2 serves a copy of a job that has received no service yet, call this job   $a$. The proof of Proposition~\ref{prop:iid_NBU_2_FCFS} relies on the fact that we can find a coupling 
where the next job that departs is job $b$ under both $\pi$ and $\pi'$. However, this coupling may not hold when there are more than two servers. For example, a departure from a server 3 can lead to a state where job $a$ leaves both systems $\pi$ and $\pi'$, but job $b$ does not. Moreover, job $b$ has received more service under policy $\pi$ than under policy $\pi'$, which is the opposite of what the coupling argument needs for the result to hold. {This is shown in the Gantt chart in Figure~\ref{fig:Prop_cas3}. At time $\tau$ job $a$ completes at server 3 under both $\pi$ and $\pi'$. At time $\tau'$, job $b$ completes at server 2 under $\pi$, but is not completely served under $\pi'$. }
\bigskip


\noindent\textbf{Proof of Corollary~\ref{prop:WNBUiid}:} We couple the service times of all jobs of class~$\{s\}$, for $s=1,2$ {for both policies}. Since the second-level policy is non-preemptive, FCFS and ROS will be equivalent for class-$\{s\}$ jobs, for $s=1,2$.

Assume that at time $t$, jobs of class $\{1,2\}$ have priority in server $s$, 
{and that $\pi$ follows FCFS for class $\{1,2\}$ at time $t$. Let job $b$ be the job of this class in server $s$ under $\pi$. Then the job in service in server $\hat s$ under $\pi$ will either be job $b$ or a job of class $\{ \hat s \}$.} 
Under policy $\pi'$ we let server $s$ serve a class-$\{1,2\}$ job chosen uniformly at random among the ones present in the queue, call it job $a$. We can have three different scenarios:

$i)$ Job $a$ and job $b$ are the same job, that is, the copy that enters service in server $s$ under policy $\pi'$ is a copy of job $b$. Then, coupling the service times of both jobs and 
{letting $\pi'$ agree with $\pi$ from time $t$ on, we have $\{N'(s)\}=\{N(s)\}$. 
}

{
$ii)$ Job $a$ and job $b$ are different jobs, and, under $\pi$ at time $t$, job $b$  has received no service in server $\hat s$. In this case, the job in server $\hat s$ under $\pi$ must be of class $\{\hat s\}$. We rename the job $a$ in $\pi'$ to be job $b$ and let $\pi'$ agree with $\pi$, and  $\{N'(s)\}=\{N(s)\}$. 
}

{
$iii)$ 
Job~$a$ and job~$b$ are different jobs, and there is a copy of job~$b$ in server~$\hat s$ that has received some service or job $b$ is starting service on server $\hat s$ under $\pi$. By repeating the second part of the argument in Proposition~\ref{prop:iid_NBU_2_FCFS}, we have that $\{N'(s)\}\leq\{N(s)\}$. 
}


\qed
\bigskip


\noindent\textbf{Proof of Proposition~\ref{prop311}:}
We first consider $\mathbb E(N^{MRF-FCFS})$. Because jobs have identical copies, we know that class-$S$ jobs will complete service in the server with highest capacity, that is, $ \mathbb E(X_{S}) =\min_{s\in S}\{1/\mu_s\}$, where $X_c$ is the service time of a type-$c$ job. Class $S$ sees a single server with capacity $\max_{s\in S}\{\mu_s\}$, and class $\{i\}$ a priority queue where class $S$ receives priority over class $\{i\}$. We assume, without loss of generality, that $\mu_1=\max_{s\in S}\{\mu_s\}$. Using \cite{HB13}, we then obtain
\begin{eqnarray}
\mathbb E(N^{MRF})  & = &  \lambda p_{S} \left( \frac{1}{\mu_1} +\frac{\lambda p_{S}\frac{1}{2 \mu_1}}{1-\lambda p_{S}\frac{1}{\mu_1}}\right)\nonumber \\
&& + \sum_{i=1}^K\frac{\lambda p_{\{i\}}}{1-\lambda p_{S}\frac{1}{\mu_1}}\ \left(\frac{1}{\mu_i}+\frac{\lambda p_{\{i\}} \frac{1}{2\mu_i} + \lambda p_{S}\frac{1}{2\mu_1}}{1-\lambda p_{S}\frac{1}{\mu_1}- \lambda p_{\{i\}}\frac{1}{\mu_i}}\right).\label{eq:prop:1}
\end{eqnarray}
For the LRF-FCFS policy, we do not have a closed-form expression for the mean number of jobs. We therefore construct an upper-bound (UB) system under the LRF-FCFS policy where a class-$S$ job departs only when the copy in server 1 is served. One can easily verify that this system upper bounds the original system. The mean number of jobs under the upper-bound system can be easily calculated, since now class $\{i\}$ sees a single server queue and class $S$ a priority queue where class 1 is given priority. Using \cite{HB13}, we then obtain 
\begin{eqnarray}
\mathbb E(N^{UB}) & = &  \sum_{i=1}^K \lambda p_{\{i\}} \left(\frac{1}{\mu_i} +\frac{\lambda p_{\{i\}} \frac{1}{2\mu_i}}{1-\lambda p_{\{i\}}\frac{1}{\mu_i}}\right)\nonumber \\
&& +  \frac{\lambda p_{S}}{1-\lambda p_{\{1\}}\frac{1}{\mu_1}} \left( \frac{1}{\mu_1} +\frac{\lambda p_{\{1\}} \frac{1}{\mu_1}+\lambda p_{S} \frac{1}{\mu_1}}{2(1-\lambda p_{\{1\}}\frac{1}{\mu_1}-\lambda p_{S} \frac{1}{\mu_1})}\right)\label{eq:prop:2}. 
\end{eqnarray}
We can now compare $\mathbb E(N^{MRF})$ with $\mathbb E(N^{UB})$. We introduce some notation to make the computations easier:
$$ a= 1-\lambda p_{S}\frac{1}{\mu_1}, \qquad b= 1-\lambda p_{\{1\}}\frac{1}{\mu_1}, \qquad c = {1-\lambda p_{S}\frac{1}{\mu_1}- \lambda p_{\{1\}}\frac{1}{\mu_1}},$$
$$\textrm{ and for }  i=1,\ldots,K, \qquad d_i = {1-\lambda p_{S}\frac{1}{\mu_1}- \lambda p_{\{i\}}\frac{1}{\mu_i}}, \qquad e_i=1- \lambda p_{\{i\}}\frac{1}{\mu_i}.$$
Let us start by comparing the performance on server $i>1$, that is, the terms multiplied by $\lambda p_{\{i\}}$, $i>1$, in Eq.~(\ref{eq:prop:1}) and Eq.~(\ref{eq:prop:2}).
\bigskip

\noindent $\displaystyle\sum_{i=2}^K [\mathbb E(N^{MRF}_i)-\mathbb E(N^{UB}_i)]$
\begin{eqnarray}
& = & \sum_{i=2}^K \left[\frac{\lambda p_{\{i\}}}{a}\left(
\frac{1}{\mu_i} + \frac{\lambda p_{\{i\}} \frac{1}{\mu_i} + \lambda p_{S} \frac{1}{\mu_1}}{2d_i}
\right) - \lambda p_{\{i\}} \left(\frac{1}{\mu_i} +\frac{\lambda p_{\{i\}} \frac{1}{\mu_i}}{2e_i}\right)\right]\nonumber\\
&= &\sum_{i=2}^K\left[\frac{\lambda p_{\{i\}}\frac{1}{\mu_i}(\lambda p_{S} \frac{1}{\mu_1})}{a} + \frac{\lambda^2 p_{\{i\}}^2  \frac{1}{\mu_i} }{2ad_ie_i}\left(\lambda p_{S}\frac{1}{\mu_1}(d_i+1)  \right) + \frac{\lambda^2 p_{\{i\}}p_{S} \frac{1}{\mu_1}}{2ad_i}\right]\nonumber\\
&\geq&\sum_{i=2}^K \left[\frac{\lambda^2 p_{\{i\}}p_{S} }{a\mu_1^2} + \frac{\lambda^2 p_{\{i\}} p_{S}}{2ad_ie_i\mu_1}\left(\lambda p_{\{i\}}\frac{1}{\mu_1}(d_i+1) + e_i\right)\right] >0.\label{eq:37_1}	
\end{eqnarray}
The last inequality holds since we have assumed that $\mu_1=\max_{s\in S}\{\mu_s\}$. 
This difference is strictly positive for any parameter values. Let us compute the difference now in server 1, that is, the terms in (\ref{eq:prop:1}) and (\ref{eq:prop:2}) multiplied by $\lambda p_{\{1\}}$ and $\lambda p_{S}$.
\bigskip

\noindent $\mathbb E(N^{MRF}_1)-\mathbb E(N^{UB}_1)$
\begin{eqnarray}
&=&  \frac{\lambda p_{\{1\}}}{a\mu_1}\left(1 + \frac{\lambda p_{\{1\}} + \lambda p_{S}}{2 c} \right) - \frac{\lambda p_{\{1\}}}{\mu_1}\left( 1 +  \frac{\lambda p_{\{1\}}}{2 b} \right)\nonumber\\
&& + \frac{\lambda p_{S}}{\mu_1}\left( 1 +  \frac{\lambda p_{S}}{2 a} \right) - \frac{\lambda p_{S}}{b\mu_1}\left(1 + \frac{\lambda p_{\{1\}}+ \lambda p_{S}}{2 c} \right)\nonumber\\
&=& \frac{\lambda^2 p_{\{1\}} p_{S}}{\mu_1^2a} + \frac{\lambda^2  p_{\{1\}}^2  }{2abc\mu_1^2}\left(\lambda p_{S}(c+1) \right) +  \frac{\lambda^2  p_{\{1\}} p_{S}}{2ac\mu_1} \nonumber\\
&&  - \frac{\lambda^2 p_{\{1\}}p_{S}}{b\mu_1^2} - \frac{\lambda^2  p_{S}^2}{2abc\mu_1^2}\left(\lambda p_{\{1\}}(c+1) \right) -  \frac{\lambda^2  p_{\{1\}} p_{S}}{2bc\mu_1} \nonumber\\
&=&  \left(p_{S} -p_{\{1\}} \right)\left( \frac{\lambda^3 p_{\{1\}}p_{S}}{ab\mu_1^3} +  \frac{\lambda^3  p_{\{1\}} p_{S}(1-c)}{2abc\mu_1^2}+  \frac{\lambda^3  p_{\{1\}} p_{S}}{2abc\mu_1^2}\right).\label{eq:37_2}
\end{eqnarray}
We note that this is nonnegative if and only if $p_{S} \geq p_{\{1\}}$. 

We can obtain a more accurate result by summing the terms multiplied by $\lambda^3$ in Equation~\eqref{eq:37_1} restricted to class $\{i\}$ and the last term in Equation~\eqref{eq:37_2}. We have then,
$$ \frac{\lambda^3p_{S}}{2a\mu_1^2}\left( \frac{p_{\{i\}}^2(d_i+1)}{d_ie_i} + \frac{p_{\{1\}}(p_{S}-p_{\{1\}})}{cb}\right).$$
We note that $a, b, c, d_i$ and $e_i$ are positive. Hence, the first multiplying term is positive. The latter implies that the whole expression is positive if and only if
$$\frac{p_{\{i\}}^2(d_i+1)}{d_ie_i} + \frac{p_{\{1\}}(p_{S}-p_{\{1\}})}{cb}>0\Longleftrightarrow $$
$$ \frac{p_{\{i\}}^2(2-\frac{\lambda p_{S}}{\mu_1}-\frac{ \lambda p_{\{i\}}}{\mu_i})}{(1-\frac{\lambda p_{S}}{\mu_1}- \frac{\lambda p_{\{i\}}}{\mu_i})(1- \frac{\lambda p_{\{i\}}}{\mu_i})} + \frac{p_{\{1\}}(p_{S}-p_{\{1\}})}{(1-\frac{\lambda p_{\{1\}}}{\mu_1})(1-\frac{\lambda p_{S}}{\mu_1}-\frac{ \lambda p_{\{1\}}}{\mu_1})}>0$$
We define $\rho_j=p_{\{j\}}/\mu_j$ for $j=1,i$ and $\rho_S=p_S/\mu_1$.
$$\Longleftrightarrow  \lambda^2 (\rho_S+\rho_i)[p_{\{i\}}^2(\rho_S+\rho_i) +p_{\{1\}}(p_{S}-p_{\{1\}})\rho_i] $$
$$-\lambda[ p_{\{i\}}^2 (2\rho_1+3\rho_S+\rho_i) +p_{\{1\}}(p_{S}-p_{\{1\}})(2\rho_i+\rho_S)]+ 2p_{\{i\}}^2+p_{\{1\}}(p_{S}-p_{\{1\}})>0.$$
For ease of notation, let
\begin{eqnarray*}
\tilde a_i & = & (\rho_S+\rho_i)[p_{\{i\}}^2(\rho_S+\rho_i) +p_{\{1\}}(p_{S}-p_{\{1\}})\rho_i],\\
\tilde b_i & = & p_{\{i\}}^2 (2\rho_1+3\rho_S+\rho_i) +p_{\{1\}}(p_{S}-p_{\{1\}})(2\rho_i+\rho_S),\\
\tilde c_i & = & 2p_{\{i\}}^2+p_{\{1\}}(p_{S}-p_{\{1\}}).
\end{eqnarray*}
Hence, if $\lambda$ is either smaller than $\lambda_{0,i}$ or larger than $\lambda^{0,i}$, then the mean number of jobs under the MRF system is strictly larger than that under UB (and hence under LRF), where
\begin{eqnarray}
\lambda_{0,i} =\frac{\tilde b_i - \sqrt{\tilde b_i^2-4\tilde a_i\tilde c_i}}{2\tilde a_i}, \textrm{ and }	\lambda^{0,i} =\frac{\tilde b_i + \sqrt{\tilde b_i^2-4\tilde a_i\tilde c_i}}{2\tilde a_i},
\end{eqnarray}
Thus, we define 
\begin{eqnarray}
\label{eq:2}\lambda_0 = \max_{i=2}^K \left\{\lambda_{0,i}\right\} = \max_{i=2}^K \left\{\frac{\tilde b_i - \sqrt{\tilde b_i^2-4\tilde a_i\tilde c_i}}{2\tilde a_i} \right\}. 
\end{eqnarray}
In particularly, if the arrival rate $\lambda$ is smaller than $\lambda_{0}$, then the mean number of jobs under the MRF system is strictly larger than that under UB (and hence under LRF). \qed
\bigskip


\noindent\textbf{Proof of Proposition~\ref{lemma1}:}
We actually show a stronger result: we prove that the system with i.i.d. copies is optimal when for each job, we can choose either independent or identical copies.

We consider a policy $\pi$ where servers are required to follow $\Pi_1$-FCFS and idling is allowed. 
Under policy $\pi$, whenever a server starts serving a first copy of a job, that server determines whether the service times of the copies of that job are identical or i.i.d.. This decision is independent of the history of the process.  
We show that for this policy $\pi$, if at time $t$ it idles or schedules identical copies for a job in service for the first time, we can construct a policy $\pi'$ with a coupled sample path, such that at time $t$ $\pi'$ does not idle or samples i.i.d. copies, and such that $\{N'(s)\}_{s\geq 0}\leq \{N(s)\}_{s\geq 0}$ with probability 1, where $N$ ($N'$) denotes the number of jobs in the system under policy $\pi$ ($\pi'$). The result follows by starting at time 0 and repeating the argument each time a policy deviates from the policy $\pi$, until we have the non-idling $\Pi_1$-FCFS policy with i.i.d. copies.

One can easily verify that the  non-idling $\Pi_1$-FCFS policy is better than idling $\Pi_1$-FCFS {for a fixed choice of independent versus identical copies}, by following the steps in the second part of the proof of Proposition~\ref{lem:NWUiid}.

Therefore let us assume that $\pi$ never idles but that at some time $t$, $\pi $ starts serving the first copy of some job, call it job 1, and $\pi$ chooses identical copies for that job. 
We let $\pi ^{\prime }$ agree with $\pi $ before time $t$ and let it choose i.i.d.\ copies for job 1. We denote by $\tau $ the time that job 1
completes service under $\pi $, say at server 1. Because the copies are
identical, server 1 has done the most work on job 1 between times $t$ and $%
\tau $. We couple the service time of job~1 on server 1 under $\pi
^{\prime }$ to that under $\pi $. Then, $\pi'$ samples i.i.d.\ copies for the rest of the copies of job 1. We let all other service times and all arrival times be coupled under both policies. We let $\pi ^{\prime }$ agree with $\pi $ until
job 1 departs under $\pi ^{\prime }$, at time $\tau ^{\prime }$. From our
coupling, $\tau ^{\prime }\leq \tau $. Let $\pi ^{\prime }$ idle any server
that serves job 1 under $\pi $ between times $\tau ^{\prime }$ and $\tau $
and let it otherwise agree with $\pi $ from time $\tau ^{\prime }$ on. From
our argument above, a policy that agrees with $\pi ^{\prime }$ but does not
idle will have even earlier departures than $\pi ^{\prime }$. The argument
can be repeated, each time reducing the number-in-system process, until
we have all i.i.d.\ copies and no idling.
\qed
\bigskip

\subsection*{B: Proofs of Section~4}
\subsubsection*{Proof of Proposition~\ref{stab:lrf} and Proposition~\ref{stab:lrf-ros}}

In order to prove both propositions, we analyze the fluid-scaled system. We recall that the redundancy structure is nested and that $N_c(t)$ denotes the number of class-$c$ jobs at time $t$. For $r>0$, we denote by $N_{c}^{r}(t)$ the system where the initial state satisfies $N^r_{c}(0)=rn_c(0)$, for all $c\in \mathcal C$. We write the fluid-scaled number of jobs per class by using standard arguments, see \cite{Bramson06},
\begin{equation}
\label{eq:frelationiid}
\frac{N_c^{r}(rt)}{r}= n_c(0) +   \frac{1}{r}\tilde A_c(rt)  - \frac{1}{r}\tilde S_{c}(T_c^{r}(rt)),
\end{equation}
where $\tilde A_c(t)$ and $\tilde S_{s}(t)$ are independent Poisson processes having rates $ \lambda p_c$ and 1, respectively. $T^{r}_{c}(t)$ is the cumulative amount of capacity spent in serving class-$c$ jobs, which strongly depends upon the correlation structure among the copies, that is, $$T^{r}_{c}(t) = g((T^{r}_{s,c}(t))_{s\in c}),$$ where $T^{r}_{s,c}(t)$ is the cumulative amount of capacity spent on serving class-$c$ jobs in server~$s\in c$ during the time interval $(0,t]$ and $g$ is characterized by the correlation structure of the copies. We note that when copies are i.i.d. $T^{r}_{c}(t)=\sum_{s\in c} T^{r}_{s,c}(t)$ and when copies are identical, $T^{r}_{c}(t)=\max_{s\in c} \{T^{r}_{s,c}(t)\}$.

In the following result, we obtain the general characterization 
of a fluid limit. The existence of fluid limits can be proved following the same steps as in \cite{Bramson06}, so the proof is omitted. 

\begin{lemma}
\label{lem:sub}
For almost all sample paths $\omega$ and sequences $r_k\to\infty$, there exists a subsequence $r_{k_j}\to\infty$ such that for all $c\in \mathcal C$ and $t\geq0$,
\begin{equation}\label{eqfluid}
\lim\limits_{j\rightarrow\infty} \frac{N_c^{r_{k_j}}(r_{k_j}t)}{r_{k_j}}=n_c(t) \ 
{\textrm{ uniformly on compact sets}}
\end{equation}
and
\begin{equation}\label{eqfluid2}
\lim\limits_{j\rightarrow\infty} \frac{T_c^{r_{k_j}}(r_{k_j}t)}{r_{k_j}}=\tau_c(t) \ {\textrm{ uniformly on compact sets}},
\end{equation}
with $(n_c(\cdot),\tau_c(\cdot))$ continuous functions.
In addition,  
\begin{equation}\label{eq:fluid}n_c(t)=n_c(0) + \lambda p_ct-\tau_c(t),\end{equation}
where $n_c(t)\geq0$, $\tau_c(0)=0$, $\tau_c(t)\leq t \max_{s\in c}\{\mu_s\}$, and $\tau_c(\cdot)$ is a non-decreasing and Lipschitz continuous function for all $c\in \mathcal C$.
\end{lemma}

In order to provide the characterization of the fluid limit, we first introduce some notation. Let us group the classes with respect to their number of copies. We denote by $\mathcal L_i$ the set of classes with $i$ copies, that is, for $i=2,\ldots, |C|$, 

$$\mathcal L_i=\{ c\in\mathcal C : |c|=i \}.$$

From  the nested structure of the system, we note that for each $c\in\mathcal L_i$ and $\tilde c\in\mathcal L_j$ with $j<i$, either $\tilde c\subset c$ or $\tilde c\cap c=\emptyset$. For all $c\in\mathcal L_i$, let us denote by $\mathcal L_i(c)=\{\tilde c \in\mathcal C :\tilde c\subset c\}$ the job classes that are subsumed in class $c$, for $i=1,\ldots,|C|$. 

We denote by $P_s(\vec N)$, the probability that, given $\vec N(0)=\vec N$, at time $t=0$ a given server~$s$ is serving a copy that is not in service in any other server. Then, the following lemma is true.

\begin{lemma}\label{lemmaROS_p} 
Consider a redundancy system with a nested topology and where servers implement $\Pi_1$-ROS. For any server $s\in S$ and $\vec N^r(0)=r\vec n^r$, such that\ $\lim\limits_{r\to\infty} \sum_{c\in\mathcal C(s)} rn_c^r >0$, then
\begin{equation}
\label{eq:1}
\lim_{r\to\infty} P_s(r\vec n^r) =  1. 
\end{equation}
\end{lemma}

\noindent\textit{Proof: }
Assume at time~0, server $s$ idles and that we are in state $\vec N^r(0)=\vec N$. At this moment, under $\Pi_1$-ROS server $s$ can only serve a single class in the system, say class $c$. Let us consider servers $l\in c$ that are serving a class-$c$ job at time $t=0$, say servers $S(c)$. We denote by $-\tilde T^r_l<0$ the time at which server $l$ started serving a new class-$c$ job, regardless of whether another job departed from the server, or the class-$c$ job preempted a copy with lower priority in that server. We note that $\frac{N_c(-\tilde T^r_l)-1}{N_c(-\tilde T^r_l)}$ is the probability that server~$l$ is \emph{not} serving the copy of the same job that is now in service in server~$s$. Hence,
\begin{equation}
\label{mo}
P_s(\vec N) = \Pi_{l\in S(c), l\neq s} \frac{N_c^r(-\tilde T^r_l)-1}{N_c^r (-\tilde T^r_l)}.
\end{equation}

We set $\vec N(0)=r\vec n^r$. Since the transition rates $\mu_s$ and $\lambda$ are of order $O(1)$, it follows directly that $\tilde T_s^r$ and $\vec N(-\tilde T_s^r)- \vec N(0)$ are of order $O(1)$ as well, so that   
\begin{equation}
\label{eq:roschoose}
\lim_{r\to\infty}\frac{N_c^r(-\tilde T_l^r)-1}{N_c^r(-\tilde T_l^r)}=\lim_{r\to\infty}\frac{N^r_c(0)-1}{N^r_c(0)}=1.
\end{equation}

It hence follows  from~\eqref{mo} that $\lim_{r\to\infty} P_s(r\vec n^r) =  1.$
\qed
\bigskip

Let us characterize the instantaneous departure rate of a class-$\tilde c$ job. Let us denote by $\mathcal C_{\tilde c}$ the classes that are a subsumed in class $\tilde c$. That is, 
$$\mathcal C_{\tilde c} = \{ c\in\mathcal C \ : \ c\subseteq \tilde c \}.$$
Note that if $\tilde c\in\mathcal L_i$, then $\mathcal C_{\tilde c}=\mathcal L_i(\tilde c)$.

\begin{lemma}\label{lemmaROS} 
Assume that jobs have exponential service times and 
\begin{itemize}
\item if copies are i.i.d. copies, assume LRF-$\Pi_2$, with LRF non-preemptive and $\Pi_2$ non-idling,
\item if copies follow some general correlation structure, assume LRF-ROS.
\end{itemize}
For each class $\tilde c\in \mathcal C$, the fluid limit $\sum_{c\in \mathcal C_{\tilde c}}n_c(t)$ satisfies the following:
$$
\frac{\mathrm{d} \sum_{c\in\mathcal C_{\tilde c}}n_c(t)}{\mathrm{d}t}= \sum_{c\in\mathcal C_{\tilde c}} \lambda p_c - \sum_{s\in \tilde c}\mu_s,  \ \ \mbox{if } \ \ n_{\tilde c}(t) > 0.
$$
\end{lemma}

\noindent\textit{Proof: } 
We first consider a general correlation structure among the copies and LRF-ROS.  When starting in state $\vec N(0)=r \vec n^r$, the drift function is
\begin{eqnarray} \label{eq:lrf_stab}
\tilde f( r \sum_{c\in\mathcal C_{\tilde c}} n_c^r)
& = & \sum_{c\in\mathcal  C_{\tilde c}}\lambda p_c\nonumber \\
&& - \sum_{s\in \tilde c}\mu_s\left(\prod_{l\in \tilde c} P_l(r \vec n^r)\right)    - g_{\tilde c}(r \vec n^r)(1-\prod_{l\in \tilde c} P_l(r \vec n^r))
\end{eqnarray}  
for $n_{\tilde c}>0$, where $ g_{\tilde c}$ is a function that captures the instantaneous departure rate of the system when more than one copy (with any correlation structure) of the same job is in service, and note that $g_{\tilde c}=O(1)$ as $r\to\infty$. Assuming that $n_{\tilde c}>0$, due to LRF and the nested structure, each server $s\in\tilde c$ s jobs of a single class $c$, with $c\subseteq\tilde c$ which is the class with least redundant copies in that server $s$. We note that the first departure term in Eq.~(\ref{eq:lrf_stab}) represents departures from servers $s\in \tilde c$ of class-$c$ jobs $c\in \tilde c$ ,  who were served in one unique server. Since these jobs have no other copies in service, the total departure rate from the servers in $\tilde c$ equals $\sum _{s\in \tilde c}  \mu_s$. The second departure term in \eqref{eq:lrf_stab} represents departures due to a class-$c$ job that is being served in more than one server simultaneously. 

Therefore, from equation (\ref{eq:lrf_stab}) together with~Lemma~\ref{lemmaROS_p}, we obtain
\begin{align}  
&\lim_{r\to\infty}	\tilde f( r \sum_{c\in\mathcal  C_{\tilde c}} n_c^r)
=  \lambda \sum_{c\in\mathcal  C_{\tilde c}} p_c - \sum_{s\in \tilde c} \mu_s.
\end{align} 

Now assume copies are i.i.d. copies and servers implement LRF-$\Pi_2$ with LRF non-preemptive and $\Pi_2$ non-idling. Under these assumptions, the drift function is given by 
\begin{align}  
&\tilde f( r \sum_{c\in\mathcal  C_{\tilde c}} n_c^r)
=  \lambda \sum_{c\in\mathcal  C_{\tilde c}} p_c - \sum_{s\in \tilde c} \mu_s, \textrm{ when } n_{\tilde c}>0 .
\end{align} 
This can be seen as follows. When there are class-$c$ jobs present, $c\in\tilde c$, each server in class $\tilde c$ is working on such a job, possibly the same job. Because of the i.i.d. copies assumption, the departure rate of these jobs is simply the sum of all the capacities. 
\qed
\bigskip

In order to prove Proposition~\ref{stab:lrf} and Proposition~\ref{stab:lrf-ros}, we introduce the redundancy degree per class. 
Let us assume w.l.o.g. that $\mathcal L_{1}$ is non-empty. For each class $c\in \mathcal L_{1}$, the fluid limit $n_c(t)$ is given by the following due to Lemma~\ref{lemmaROS}: 
$$
\frac{\mathrm{d} n_c(t)}{\mathrm{d}t}= \lambda p_c - \sum_{s\in c}\mu_s,  \ \ \mbox{if } \ \ n_{c}(t) > 0.
$$
We note that $\lambda p_c - \sum_{s\in c}\mu_s<0$, by assumption. This coincides with the fluid limit of an $M/M/1$ system with arrival rate $\lambda p_{c}$ and server capacity $\sum_{s\in c}\mu_s$. Hence, for all $c\in\mathcal L_1$, $n_c(t)$ reaches zero in finite time, say at time $T_1$, and stays zero. 

The proof follows now by induction. Assume that there is a time $T_j$, such that $n_c(t)=0$ for $t>T_j$, for all $c\in\mathcal L_i$ and $i\leq j$. In the following we show that for $\mathcal L_{j+1}$, there is a $T_{j+1}>T_j$, such that $n_c(t)=0$ for $t>T_{j+1}$ and for all $c\in \mathcal L_{j+1}$. 

For a class $c\in\mathcal L_{j+1}$, the fluid drift of $\sum_{\tilde c\in\mathcal L_{j+1}(c)} n_{\tilde c}(t)$ is given by the following due to Lemma~\ref{lemmaROS}: 
\begin{equation}\label{lrf-ros:eq:2}
\frac{\mathrm{d} \sum_{\tilde c\in\mathcal L_{j+1}(c)}n_{\tilde c}(t)}{\mathrm{d}t}= \sum_{\tilde c\in\mathcal L_{j+1}(c)} \lambda p_{\tilde c} - \sum_{s\in c}\mu_s,  \ \ \mbox{if } \ \ n_{c}(t) > 0.
\end{equation}

From the induction hypothesis, we note that there exits time $T_j$ such that $n_{\tilde c}(t)=0$ for all $\tilde c\in\mathcal L_i$ with $i\leq j$. Hence, $
\frac{\mathrm{d} \sum_{\tilde c\in\mathcal L_{j+1}(c)}n_{\tilde c}(t)}{\mathrm{d}t}=\frac{\mathrm{d} n_{c}(t)}{\mathrm{d}t}$, for $t\geq T_{j}$. Together with Eq.~\ref{lrf-ros:eq:2},

$$
\frac{\mathrm{d} n_{c}(t)}{\mathrm{d}t}= \sum_{\tilde c\in\mathcal L_{j+1}(c)} \lambda p_{\tilde c} - \sum_{s\in c}\mu_s,  \ \ \mbox{if } \ \ n_{c}(t) > 0,
$$
for all $t\geq T_j$. We note that $\lambda \sum_{\tilde c\in\mathcal L_{j+1}(c)} p_{\tilde c} - \sum_{s\in c}\mu_s<0$, by assumption. This coincides with the fluid limit of an $M/M/1$ system with arrival rate $\lambda \sum_{\tilde c\in\mathcal L_{j+1}(c)} p_{\tilde c}$ and server capacity $\sum_{s\in c}\mu_s$. Hence, for all $c\in\mathcal L_{j+1}$, $n_c(t)$ reaches zero in finite time, say at time $T_{j+1}$, and stays zero. Hence, there exists time $\tilde T>0$ when the fluid process is empty. Together with \cite{Robert03}, we conclude that the system is stable.
\qed
\bigskip 

\end{document}